\documentclass{aa}  
\usepackage{graphicx}
\usepackage{txfonts}
\usepackage{physics}

\newcommand{\noi}{\noindent}

\usepackage{color}

\begin{document} 

\title{Nonlinear diffusion of cosmic rays escaping from supernova remnants: Cold partially neutral atomic and molecular phases}
  \titlerunning{Nonlinear diffusion of cosmic rays escaping from supernova remnants}

   \author{L. Brahimi \inst{1},  A.Marcowith \inst{1}
          \and
          V.S. Ptuskin \inst{2}
          }

   \institute{Laboratoire Univers et Particules de Montpellier (LUPM) Universit{\'e} Montpellier, CNRS/IN2P3, CC72, place Eug{\`e}ne Bataillon,\\ 34095, Montpellier Cedex 5, France.\\
              \email{Loann.Brahimi@umontpellier.fr}
         \and
             Pushkov Institute of Terrestrial Magnetism, Ionosphere and Radiowave Propagation, 108840, Troitsk, Moscow, Russia.\\
             }

   \date{Received -; accepted -}

 
  \abstract
  {}
{We aim to elucidate cosmic ray (CR) propagation in the weakly ionized environments of supernova remnants (SNRs) basing our analysis on the cosmic ray cloud (CRC) model developed by \citet{2013ApJ...768...73M} and \citet{2016MNRAS.461.3552N}.}  
   {We solved two transport equations simultaneously: one for the CR pressure and one for the Alfv\'en wave energy density where CRs are initially confined in the SNR shock. Cosmic rays trigger a streaming instability and produce slab-type resonant Alfv\'en modes. The self-generated turbulence is damped by ion-neutral collisions and by noncorrelated interaction with Alfv\'en modes generated at large scales.}
   {We show that CRs leaking in cold dense phases such as those found in cold neutral medium (CNM) and diffuse molecular medium (DiM) can still be confined over distances of a few tens of parsecs from the CRC center for a few thousand years. At 10 TeV, CR diffusion can be suppressed by two or three orders of magnitude. This effect results from a reduced ion-neutral collision damping in the decoupled regime. We calculate the grammage of CRs in these environments. We find that  in both single and multi-phase setups at 10 GeV, CNM and DiM media can produce grammage in the range 10-20 $\rm{g/cm^2}$ in the CNM and DiM phases. At 10 TeV, because of nonlinear propagation the grammage increases to values in the range 0.5-20 $\rm{g/cm^2}$ in these two phases. We also present preliminary calculations in inhomogeneous interstellar medium combining two or three different phases where we obtain the same trends.}

 {}


   \maketitle
%

\section{Introduction}\label{S:Intro}

Cosmic rays (CRs) are a major component of the interstellar medium (ISM), along with the gas and magnetic field \citep{2001RvMP...73.1031F}, and have a non-negligible impact on the global dynamics of our galaxy \citep{2015ARA&A..53..199G}. Through their streaming they can drive galactic winds \citep{2017MNRAS.470..865R, 2016ApJ...816L..19G}. They contribute to the enrichment of the ISM in light elements by spallation and are also a strong source of ionization especially in the dense and cold ISM \citep{2009A&A...501..619P}. Furthermore, they may contribute to the turbulent dynamics of our galaxy through the generation of magnetic perturbations \citep{Commercon19}.

The origin of CRs is still under debate \citep{2001SSRv...99..329D}. Strong shocks in supernova remnants (SNRs) seem to be able to accelerate CRs via diffusive shock acceleration (DSA) up to a few hundred TeV \citep{1983A&A...125..249L} or even to PeV at early times (when shock speeds $v_{\rm sh}$ are as high as 0.1 times the speed of light, e.g. \citet{2013MNRAS.435.1174S}). Once released into the ISM, CRs have a random walk induced by scattering off turbulent magnetic perturbations generated by large-scale plasma motions and/or by the waves that they self-generate while drifting at speeds greater than the local Alfv\'en speed \citep{1968ApJ...152..987W, 1969ApJ...156..445K, 1975MNRAS.172..557S, 2004ApJ...614..757Y, 2004ApJ...604..671F}.  However, currently there is no fully consistent theory capable of simultaneously explaining both CR acceleration at SNR shocks and their escape into the ISM (see \citet{2012A&A...541A.153T} for a discussion). Besides ionization studies, probes of ISM and CR interactions are essentially based on gamma-ray observations, such as for example the recent H.E.S.S. Tcherenkov telescope observations from the forward shock of the SNR RX J 1713.7-3946 \citep{2018A&A...612A...6H} or W 28 \citep{2008A&A...481..401A} which provide new constraints on the CR escape process.\\
    
The ISM is a very complex medium and the way it is structured is still an open question. It is accepted that the ISM can be modeled in terms of distinct phases in rough pressure equilibrium. Depending on the degree of ionization, we find first the ionized phases: the hot ionized medium (HIM) and the warm ionized medium (WIM), and then the partially ionized phases: the warm neutral medium (WNM), the cold neutral medium (CNM), the diffuse molecular medium (DiM), and molecular clouds (MCs); see e.g. \citep{1969BAAS....1..240F, 1977ApJ...218..148M, 1989ApJ...345..811R, 2008PASA...25..184G}. Partially ionized phases cover a minor fraction of the galactic disk volume but contain the majority of the mass. Most of the ISM matter is concentrated in MCs, the birthplace of stars. In partially ionized medium, low-energy CRs constitute an important source of ionization \citep{2018MNRAS.480.5167P}\footnote{Energetic particles with a kinetic energy in the range between MeV and 100 MeV for protons and 100 eV and keV for electrons \citep{2009A&A...501..619P}.}, especially close to an accelerator where one can expect enhanced CR density for some amount of time \citep{2014A&A...568A..50V, 2016MNRAS.461.3552N}. Cosmic rays can also exert a force on the gas through the production of magnetohydrodynamic (MHD) waves. The turbulent dynamics of WNM and CNM as well as MCs may be affected in the presence of an intense CR flux (\citet{2011ApJ...739...60E}, \citet{Commercon19}). \\ 
    
 In this paper, we particularly focus on the way CRs escape from SNR shocks into the ISM, the way they interact with the partially ionized ISM phases, and how their transport properties are modified in these media. Our model relies on the CR cloud model (CRC; see section \ref{SS:CRs} for further details) proposed by \citet{2013ApJ...768...73M, 2008AdSpR..42..486P} and in \citet{2016MNRAS.461.3552N}, referred to hereafter as N16. This study completes a parallel investigation of CR escape in the ionized ISM by \citet{2019MNRAS.484.2684N}, referred to hereafter as N19. Cosmic rays at different energies escape from the remnant at different times depending on the properties of the ambient ISM. In this study we discuss three different escape models when radiative losses become important during SNR expansion. Depending on the local CR over pressure, CRs trigger a resonant streaming instability and produce magnetic fluctuations \citep{1971ApJ...170..265S, 1975MNRAS.172..557S} \footnote{We restrict our investigation to the case of the generation of resonant slab Alfv\'en waves, and defer to a future study the case of the generation of other types of kinetic or convective instabilities; see however \citet{Inoue19}.}. We assume in this study that the amplitude of the self-generated waves is weak and treat the problem according to the quasi-linear theory of CR transport \citep{2002cra..book.....S}. In partially ionized phases, the waves triggered by CRs are damped by ion-neutral collisions and by nonlinear interactions with the large-scale turbulence (see N16 and \citet{2004ApJ...604..671F}). This work provides some answers to different open questions: For how long/far does a SNR produce CR over-density in the neutral/molecular ISM? What is the space-time-dependent rate of self-generated waves around a SNR accelerator? How can we evaluate the impact of CR self-confinement over the grammage (see also \citet{2016PhRvD..94h3003D})? Over what distance should one account for the effect of stronger local CR gradients? 

In section \ref{S:MOD} we present our model: we describe the properties of neutral ISM phases adopted in this study, we discuss the different wave-damping processes, we discuss the way CRs escape, and we generalize the work proposed by \citet{2013ApJ...768...73M} and N16. In particular we propose variants of the escape model based on the work of \citet{2017ApJS..229...34S}. In section \ref{S:RES} we present our results for each escape model for one particular medium at different energies. In section \ref{S:GRA} we calculate the grammage of CRs produced during their propagation close to their sources also considering a setup including multiple phases. Finally, we conclude in section \ref{S:COC}.  

\section{The model}\label{S:MOD}
        \subsection{Partially ionized interstellar medium phases}
                \subsubsection{Phase properties}
            
            We first describe the main physical properties of weakly ionized ISM phases. We start from the assumption that the weakly ionized ISM is composed of hydrogen and helium with a density ratio $n_{\rm H} / n_{\rm He} \approx$ 13.28. The total plasma density is given by $n_{\rm T} = n_{\rm H} + n_{\rm He}$. For each phase, we consider a thermal plasma using a two-fluid MHD model: each fluid (neutral and ionized) is composed of a dominant species described by its mean mass ($\mu_{\rm i} = m_{\rm i}/m_{\rm H}$ and $\mu_n= m_{\rm n}/m_{\rm H}$) and its density ($n_{\rm n}$ and $n_{\rm i}$). We define the ionization fraction $X= n_{\rm i}/(n_{\rm i}+ n_{\rm n})$ of the plasma. We consider the plasma to be filled with a magnetic field composed of a regular and a turbulent component. The turbulent component results from an injection of free energy at large scales $L \simeq 50-100~\rm{pc}$ (see N16). The essential chemical and physical properties of cold ISM have been taken from the work of \citet{2006ARA&A..44..367S} and are summarized in Table \ref{T:Phases}. These values are subject to some uncertainties that are also included in the table.\\
           
           \paragraph*{Neutral atomic phases} The WNM and CNM are thermally stable solutions of the thermal instability together with the WIM \citep{1969BAAS....1..240F}. Both phases are composed of atomic gas. The WNM and CNM are detected by HI emission and absorption line surveys. For both WNM and CNM, we assume that the dominant neutral species is atomic hydrogen (HI) while the dominant ion is ionized hydrogen (HII) in the WNM and ionized carbon (CII) in CNM, respectively \citep{2006ARA&A..44..367S}. The WNM is a relatively low density medium ($n_{\rm T} = 0.2-0.5~ \rm{cm}^3$) with a high equilibrium kinetic temperature ($6~000 - 10~ 000~\rm{K}$) whereas CNM is denser ($n_{\rm T} = 20-50~\rm{cm}^{-3}$) and colder ($T = 30-100~\rm{K}$). The typical size of the phases are difficult to derive: 21 cm line surveys give empirical relations between the gas velocity dispersion $\sigma$ and the typical size $R$ of the region, suggestive of a turbulent cascade with a scaling $\sigma \propto R^k$, with $k \sim 0.35$ \citep{1979MNRAS.186..479L, 2008MNRAS.387L..18R}. These analyses reconstruct $\sigma$ from column density and line spin temperature measurements and have $R$ ranging between $10^{-2}$ and 100 pc. Another aspect is that these media are not isotropic in shape but rather show elongated structures \citep{2003ApJ...586.1067H}. The aspect ratio between parallel and perpendicular length scales can be up to two orders of magnitude. We then decide to keep a maximum length scale for these two phases of $\sim 100$ pc which is close to the typical size of large-scale turbulent motion injection. We fix the phase size interval to $R \sim 1-100 ~ \rm{pc}$, smaller scales being of little interest in the framework of this study which involves CRC with sizes larger than 1 pc.

           \paragraph*{Diffuse molecular medium} A fundamental difference between cold atomic and cold molecular phases is the level of penetrating UV radiation \citep{2006ARA&A..44..367S}. As UV radiation decomposes molecules, molecular clouds need to be surrounded by a shielding medium in order to conserve their chemical equilibrium. This shield ensures a transition from an external part composed of atomic gas (HI dominated) to a molecular interior (H$_2$ dominated) and has a total minimal visual extinction magnitude of 0.2. For this medium, we assume a neutral fluid composed of a mixture of 50\% neutral hydrogen (HI) and 50\% molecular hydrogen (H$_2$). As enough interstellar radiation is present to photo-dissociate CO molecules, the dominant ionized species is CII as in the CNM phase. The DiM is also characterized by higher density ($n_T = 100-500 ~ \rm{cm}^{-3}$); see \citet{2006ARA&A..44..367S}. Its characteristic size is not well constrained; we consider typical sizes derived by \citet{1987ApJ...319..730S} from CO surveys between $1~\rm{pc}$ to $50~\rm{pc,}$ the upper limit corresponding to giant molecular clouds.\\
            
          We note that the properties of the phases and the dimension of our model constrain the number of cases we can study in our work. We use a model of CR propagation restricted to the propagation of CRs along a background magnetic field, hence a 1D model. This is the {\it flux tube approximation} (see section \ref{SS:CRs}), which imposes that the respective sizes of the phase and the CRC be $R_{\rm phase} > R_{\rm CRC}(t_{\rm esc}),$ where $t_{\rm esc}$ is the time at which CRs at a particular energy escape from the CRC. We verify that this condition applies a fortiori to each phase in our study. Another aspect is that the partially ionized ISM is likely a relatively inhomogeneous medium, and we should not expect the SNR to propagate into a single phase over several tens of parsecs. To account for this inhomogeneity we need to investigate CR propagation in multi-phase ISM, which can require  multi-dimensional simulations. This aspect is discussed in section \ref{S:COC} and merits future dedicated study. However, in section \ref{S:GRA} we consider an example of multi-phase ISM in the context of the calculation of the CR grammage. 

            \begin{table*} \label{T:Phases} \centering
                \begin{tabular}{cccc}
                Phase       &                 WNM            &      CNM      &    DiM     \\ 
                \hline
                \hline
                $T$ [K]     & $6\times 10^3 - 1 \times 10^4$ &    $30-100$   &  $30-100$  \\
                            &      ($8 \times 10^3$)         &     ($50$)    &   ($50$)   \\ 
                \hline
                $B_0$ [$\mu$G] & $5$                           & $6$           & $6$           \\
                \hline
                $n_{\rm T}$ [cm$^{-3}$] & $0.2-0.5$                & $20-50$       & $100-500$  \\ 
                                  & ($0.35$)                 & ($30$)        & ($300$)    \\ 
                \hline
                $X$          & $0.007-0.05$   & $4 \times 10^{-4} - 10^{-3}$ & $10^{-4}$  \\ 
                             &  ($0.02$)      & ($8 \times 10^{-4}$)         &                  \\ 
                \hline
                $n_{\rm i}$ [cm$^{-3}$] & $7 \times 10^{-3}$ & $2.4 \times 10^{-2}$& $3 \times 10^{-2}$ \\ 
                \hline
                \hline
                Neutral      & 93\% HI + 7\% He & 93\% HI + 7\% He & 93\% (0.5HI + 0.5H$_2$) + 7\% He \\ 
                $\mu_{\rm n}$ [$m_p$] & $1.21$ & $1.21$ & $1.67$ \\ 
                \hline
                Ion          & H$^+$ & C$^+$ & C$^+$ \\ 
                $\mu_{\rm i}$ [$m_p$] & $1$ & $12$ & $12$ \\ 
                \hline
                \hline
                $R$ [pc] & $1-100$ & $1-100$ & $1-50$ \\                  
                
                \end{tabular}
                \caption{Fiducial physical quantities for the WNM, CNM, and DiM phases adopted in this work. $T$ is the gas temperature, $B_0$ is the mean magnetic field strength, $n_{\rm T} = n_{\rm n} + n_{\rm i}$ is the total gas density, $X$ is the gas ionization fraction, $\mu_{\rm i}$ ($\mu_{\rm n}$) is the mean ion (neutral) mass; see N16, \citet{2009A&A...508.1099J}, \citet{2006ARA&A..44..367S} and \citet{2005ApJ...628..260N}. Values between brackets are the mean values adopted for numerical application. See the text for the justification of the selected values of the phase size R.}
            \end{table*}
  
                \subsubsection{Magnetohydrodynamic wave damping in partially ionized media}\label{S:DAM}
        
        \paragraph*{Ion-neutral damping} Alfv\'en waves are supported by the motion of ions. In a two-fluid model energy transfer rate operates between neutrals and ions which produces a damping of the Alfv\'en waves. In the WNM phase, the ion-neutral collision rate $\nu_{\rm in}$ expresses the energy transfer rate from ions to neutrals and is given in Eq. (19) of N16 by 
        
        \begin{equation} 
                \nu_{\rm in} = n_{\rm n} \sigma v_{\rm th} = 2 n_{\rm n} (8.4 \times 10^{-9} ~\rm{cm}^{-3}~\rm{s}^{-1} ) \left( \frac{T}{10^4~\rm{K}} \right)^{0.4} \ ,
        \end{equation}
        
        \noi where $\sigma v_{\rm th}$ is the fractional rate of change of the proton velocity $v_{\rm th}$ averaged over the velocity distribution, $n_{\rm n}$ ($n_{\rm i}$) is the neutral (ion) volume density in cm$^{-3}$, and $T$ is the plasma temperature in K. \\
        In the case of a colder phase ($T \le 50$ K), such as in CNM or DiM, the ion-neutral collision rate is expressed by \citep{2009A&A...508.1099J}
        
        \begin{equation}
            \nu_{\rm in} = n_n (2.1 \times 10^{-9} ~ \rm{cm}^{-3}~\rm{s}^{-1} ) \ .
        \end{equation}
        
        The neutral-ion collision rate $\nu_{\rm ni}$ expresses the energy transfer from neutrals to ions; it is linked with $\nu_{\rm in}$ by the relation \citep{2015ApJ...810...44X}
        
        \begin{equation}
            \label{eq:in_coll}
                \nu_{\rm in} = \chi \nu_{\rm ni} \ ,
        \end{equation}
        
        \noi where $\chi = (m_n/m_i) ( X^{-1} - 1) = \rho_n / \rho_i$ is the ratio between the mass density of neutrals and ions. 
        Ion-ion $\nu_{\rm ii}$ and neutral-neutral $\nu_{\rm nn}$ collision rates express the internal energy exchanges within each fluid. In a weakly ionized plasma, $\nu_{\rm ii}$ is negligible because of the low ion relative density.  In the case of very low ionization rates, $\nu_{\rm nn}$ may contribute to the damping of Alfv\'en waves (\citet{2015ApJ...810...44X}). It is however found to be negligible in the cases under study here.
        
\begin{figure} \centering 
                \includegraphics[width=0.47\textwidth]{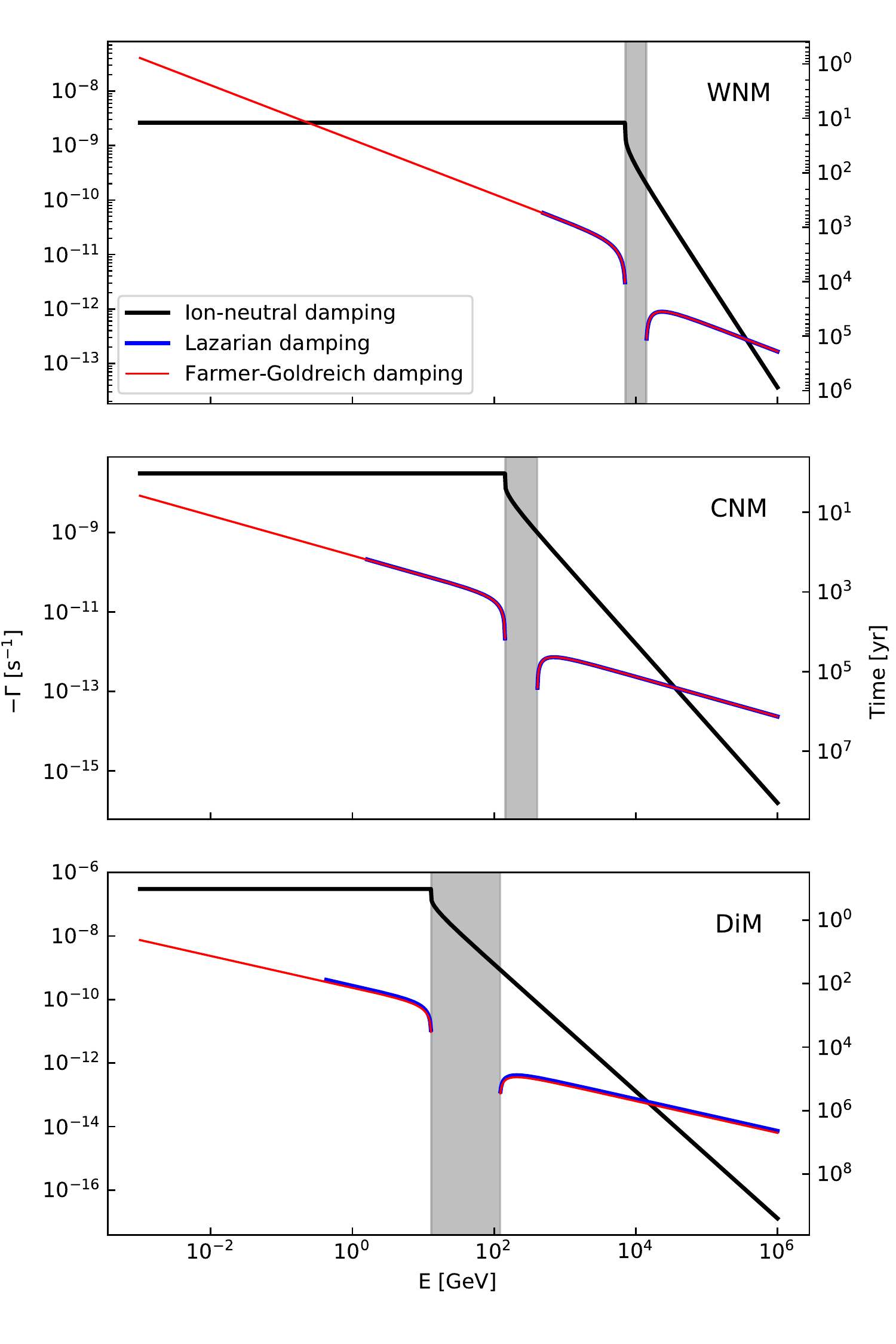}
                \caption{Ion-neutral damping rate ($\Gamma_{\rm d}^{\rm in}$, black curves), Farmer and Goldreich damping rate ($\Gamma_{\rm d}^{\rm FG}$, red curves), and Lazarian damping rate ($\Gamma_{\rm d}^{\rm L}$, blue curves) as a function of resonant CR energy. The shaded region refers to the range of energy where the real pulsation of Alfv\'en waves is not defined, which is why the turbulent damping, which depends on the Alfv\'en, is not defined in this scale range.}
                \label{fig:damping}
            \end{figure}

        As shown by Eq. (\ref{eq:in_coll}), the quantity of energy transferred between ions and neutrals depends on the ion/neutral plasma mass density. If $X \ll 1$, then $\nu_{in} \gg \nu_{ni}$ controls the energy transfer between ions and neutrals. The Alfv\'en wave dispersion relation is calculated in a two-fluid approach by \citet{2013ApJS..209...16S} and has been solved using different approximations by \citet{2016ApJ...826..166X} and \citet{2004ApJ...603..180L}. The imaginary part of the solution of the dispersion relation corresponds to the Alfv\'en wave damping rate; it is presented in figure \ref{fig:damping} for each phase by a bold black line as a function of the energy of CRs. As in this work we only consider CR in resonance with waves, there is a one to one relationship between the particle Larmor radius $r_{\rm L}= \gamma m c/qB_0$ (or energy, in the relativistic domain) and the wave number k, such that $kr_{\rm L} \sim 1$. Here we introduce the CR charge q, mass m, and Lorentz factor $\gamma=(1-(v/c)^2)^{-1/2}$ where $v/c$ is the ratio of the speed of the particle to the speed of light. The background magnetic field has a strength $B_0$. We observe two asymptotic behaviors. At low CR energy ($E \ll eB_0 V_{\rm Ai}/\nu_{\rm in}$), ions and neutrals are weakly coupled and the energy transfer from ions to neutrals is maximal because their motions are uncorrelated. The asymptotic Alfv\'en wave damping rate is then given by 
        
        \begin{equation}
                \Gamma_{\rm in} = - \frac{\nu_{\rm in}}{2} \ ,
        \end{equation}
        
        \noi and the Alfv\'en speed is the one produced by the ionic component defined by $V_{\rm Ai} = B_0 / \sqrt{4 \pi \rho_i}$. At high CR energy ($E \gg eB_0 V_{\rm A} \chi / \nu_{\rm in}$), the  motions of ions and neutrals become correlated and the energy transfer from ions to neutrals drops \citep{1977MNRAS.178...85M}. The Alfv\'en waves damping rate is therefore given by 
        
        \begin{equation}
                \Gamma_{\rm in} = - \frac{\xi_{\rm n} V_{\rm A}^2 e^2 B_0^2}{2 \chi^{-1} \nu_{\rm in}} E^{-2} \ ,
        \end{equation}
        
        \noi where $\xi_{\rm n} = \rho_{\rm n} / (\rho_{\rm n} + \rho_{\rm i})$ is the neutral fraction in the gas, and the Alfv\'en speed is given by $V_{\rm A} = B_0 / \sqrt{\rho_{\rm n} + \rho_{\rm i}}$. Finally, in the energy range $[eB_0 V_{\rm Ai}/\nu_{\rm in}, eB_0 V_{\rm A} \chi / \nu_{\rm in}]$ Alfv\'en waves do not propagate.
        
  In Fig. \ref{fig:damping} we see that the damping of Alfv\'en waves is stronger in the diffuse molecular phases (DiM) than in atomic phases (WNM, CNM). The cutoff band in the WNM is located around 10 TeV while in the other mediums it is located around 100 GeV. We can also observe that the forbidden propagation band gets wider as the ionization rate decreases.
        
        \paragraph*{Turbulent damping} Alfv\'en waves generated by CRs can also interact with large-scale injected Alfv\'enic turbulence. Although the sources of this turbulence are poorly known, among the main contributors we can cite supernovae explosions or galactic differential rotation \citep{2004RvMP...76..125M}. Interactions between wave packets propagating in opposite directions lead to a distortion of the wave packets. A CR-generated wave of wavelength $\lambda \sim r_{\rm L}$ is "damped" when the size of the distortion produced by the interacting wave packet is $\delta \sim \lambda$ (see section 4.2 in L16). In this process, waves generated by CRs interact with background perturbations through a three-wave process \citep{1997PhPl....4..605N} which leads to the production of larger k modes. This process is called turbulent damping (see \citet{2004ApJ...604..671F} (FG04),  \citet{2016ApJ...833..131L} (L16)) and we use this term hereafter. It should be stressed that this process does not correspond to a real damping but produces a transfer of perturbations from the resonant scale to smaller scales.
   
 In N16  we use the turbulent damping term proposed by FG04. The authors of L16 propose an extension of the FG04 model to other turbulent regimes, including magnetosonic turbulence. Their results are based on a particular approach to the phenomenon of MHD turbulence (see \citet{1999ApJ...517..700L}). By adopting this model we acknowledge its limitation. More involved discussions of wave interaction in either super- of sub-Alfv\'enic turbulence can be found in \citet{1994JGR....9919267M, 1996JGR...10117093Z,2012ApJ...745...35Z, 2015ApJ...805...63A}, albeit mostly in the context of solar wind turbulence. Adapted to ISM, these models may produce a different turbulent damping rate which could modify the generality of our conclusions especially at high CR energies where ions and neutrals are in the coupled regime of collisions. However, the turbulent damping rate has to increase to at least one order of magnitude to be competitive at 10 TeV (see figure \ref{fig:damping} below). \\
 
 We now reproduce the main results obtained by L16. In the sub-Alfvenic regime (with $M_{\rm A} = V_{\rm L} / V_{\rm A} < 1$) the turbulence at large scales is weak mostly 2D; it cascades perpendicularly to the mean magnetic field direction. The velocity perturbations scale as $v_{\rm k} = V_{\rm L} (kL)^{-1/2}$. Below the scale $\ell = L M_{\rm a}^2$ the turbulence becomes strong and is described $v_{\rm k} = V_{\rm L} (kL)^{-1/3} M_{\rm A}^{4/3}$. In the above expression, $M_{\rm A} = V_{\rm L} / V_{\rm A}$ is the Alfv\'enic Mach number, $V_{\rm L}$ is the rms turbulence speed at the injection scale $L$, and  $k\equiv k_\perp$ is the perpendicular wave number. \\
In the sub-Alfvenic, strong turbulence regime, ($M_{\rm A} < 1$ and $k^{-1} < LM_A^4$) the damping term is given by 
        \begin{equation}
                \Gamma_{\rm L, sub, A, s} \approx \frac{V_{\rm A} M_{\rm A}^2}{k^{-1/2} L^{1/2}} ~ \rm{for} ~ 
            \frac{l_{\rm min}^{4/3} M_{\rm A}^{4/3}}{L^{1/3}} < k^{-1} < L M_{\rm A}^4 
        ,\end{equation}
        
        \noi where $M_{\rm A} = V_{\rm L} / V_{\rm A}$ is the Alfv\'enic Mach number, $V_{\rm L}$ is the rms turbulence speed at the injection scale $L$, and $k$ corresponds to the turbulent scale. Here, $l_{\rm min}$ corresponds to the turbulence cutting scale and is given by the equation (\ref{eq:l_min}) below.
       
        In a sub-Alfvenic, weak turbulence regime ($M_{\rm A} < 1$ and $k^{-1} > L M_{\rm A}^4$) the damping term is given by 
        
        \begin{equation}
                \Gamma_{\rm L,sub,A,w} \approx \frac{V_{\rm A} M_{\rm A}^{8/3}}{k^{-2/3}L^{1/3}}
        ,\end{equation}
        
        \noi for $LM_{\rm A}^4 < k^{-1} < LM_{\rm A}$. Alternatively, if $k^{-1} > LM_{\rm A}$, then it is given by  
        
        \begin{equation}
                \Gamma_{\rm L,sub,A,w} \approx M_{\rm A}^2 \frac{V_{\rm A}}{L} ~ 
            \rm{for} ~ k^{-1} < L
        ,\end{equation}
        
        \noi and
        
         \begin{equation}
                \Gamma_{\rm L,sub,A,w} \approx M_{\rm A}^2 \frac{V_{\rm A} L}{k^{-2}}  ~ 
            \rm{for} ~ k^{-1} \gg L.
        \end{equation}
        In the super-Alfvenic turbulence regime ($M_{\rm A} > 1$), the turbulence at scales larger than $\ell=L/M_{\rm a}^3$ is isotropic and hydrodynamic and is described by a Kolmogorov model. Below this scale, strong turbulence applies and the cascade leads to anisotropic velocity perturbations, again following $v_{\rm k} = V_{\rm L} (kL)^{-1/3} M_{\rm A}^{4/3}$. The damping rate is given by 
        \begin{equation}
                \Gamma_{\rm L,super} \approx \frac{V_{\rm A} M_{\rm A}^{3/2}}{L^{1/2} k^{-1/2}} 
        \end{equation}
        
        \noi if $\frac{l_{\rm min}^{4/3}}{L^{1/3}}M_{\rm A} < k^{-1} < LM_{\rm A}^{-3}$ , and 
        
        \begin{equation}
                \Gamma_{\rm L,super} \approx \frac{V_{\rm A} M_{\rm A}}{L^{1/3} k^{-2/3}}
        \end{equation}
        
        \noi if $k^{-1} > LM_{\rm A}^{-3}$. 
        
     Although not trivial, the choice of the injection scale $L$ and the associated turbulent velocity $V_L$ is essential to characterize the nature of the turbulence and the behavior of the turbulent damping. Following the same assumptions as N16 , we chose $L = 50~\rm{pc}$ and $V_{\rm L} = \rm{max}(V_{\rm A}, c_{\rm s})$ where $V_{\rm A}$ corresponds to the Alfv\'en speed in the coupled regime and $c_{\rm s}$ is the sound speed. For each medium, this assumption leads to $M_{\rm A} \approx 1$ because $V_A$ exceeds $c_{\rm s}$. The MHD turbulence is in the trans-Alfv\'enic turbulent regime. However, in the DiM, values extracted from Table \ref{T:Phases} lead to $M_{\rm A} \sim 5$. Increasing by a factor five however does not change the impact of turbulent damping as we limit our investigation to CRs with energies below 30 TeV. Even at these energies, ion-neutral collisions dominate the damping of slab waves. Acknowledging the uncertainties over the turbulence injection scale and the magnetic field strength and densities in these media, we decided to keep a trans-Alfv\'enic turbulence injection model in all of the phases under study. The choice for the injection scale $L$ is rather arbitrary as very little is known about its value in the ISM. It is also very likely that  $L$  varies depending on the position in the disk and between the disk and the galactic halo. Typical scales of 1-100 pc are reported in \citet{2008ApJ...680..362H}. Furthermore, several sources of turbulent motion injection may compete at a given location, as may be the case in the local ISM \citep{2015ApJ...804L..31B}. Turbulence can be injected by supernovae, massive star winds, galactic shear motions, HII region expansion, or by CR themselves; see eg \citet{2017ApJ...835..258G}. 
        
        Figure \ref{fig:damping} shows the two turbulent damping models. The red solid line represents the FG04 model and the bold blue line the L16 model. In a trans-Alfv\'enic regime, the turbulent damping rate is given by 
        
        \begin{equation}
                \Gamma_{\rm{trans, L04}} \approx \frac{V_{\rm A}}{k^{-1/2}L^{1/2}} ~ \rm{for} ~ 
            \frac{l_{\rm min}^{4/3}}{L^{1/3}} < k^{-1} < L. 
        \end{equation}
        
        \noi The difference between the FG04 model and the L16 model appears at 
        
        \begin{equation}
        \label{eq:l_min}
                l_{\rm min}^{-1} = (2 \nu_{ni})^{3/2} L^{1/2} V_{\rm A}^{-3/2} \sqrt{1 + \frac{V_{\rm A}}{2 \nu_{\rm{ni}} L}}
        ,\end{equation}
        
        \noi corresponding to the low-energy cutting scale of the turbulence. Unlike in the FG04 model, the turbulent cascade in the L16 model is cut at energies below $\sim 500$ GeV, $1.6$ GeV, and $0.4$ GeV in the WNM, CNM, and DiM phases, respectively. However, in practice, this difference is not relevant in our study because at low energy, the Alfv\'en wave damping rate is always dominated by ion-neutral collisions, and the turbulent damping only becomes dominant at high energies ($\sim 100~\rm{TeV}$). 
        
        \subsection{Escape of cosmic rays  from supernova remnants} \label{SS:CRs}
    The process of CR leakage from SNRs is connected to the acceleration process in the source interior. As CRs leak out, they  extract some energy flux which increases the compressibility of the shocked fluid \citep{1999ApJ...526..385B}. High-energy particles streaming ahead of the shock front produce magnetic fluctuations which ensure the confinement of lower energies at the shock \citep{2013MNRAS.431..415B}. Until now however, no model has been built that describes in a self-consistent way the acceleration mechanism at the SNR strong shocks and the process of CR escape from the accelerator (see the discussion in \citet{2012A&A...541A.153T}), so the transition between the following two locations.   
    \begin{itemize}
        \item The source interior where the magnetic perturbations are generated by streaming instabilities and the level of turbulence is high ($\delta B \sim B_0$). Cosmic rays have a Bohm diffusion and their mean free path is small ($\lambda \sim r_{\rm L}$), meaning that they likely diffuse isotropically ($D_\parallel \sim D_\perp$). The CR pressure is in excess compared to the magnetic pressure. 
        \item The surrounding ISM where magnetic perturbations are injected at large scale ($\delta B \ll B_0$). The mean free path of CRs is high ($\lambda \gg r_{\rm L}$) and they diffuse along the mean magnetic field ($D_\parallel \gg D_\perp$). Their pressure is in equipartition with the magnetic pressure. 
    \end{itemize}
    
    \noi Current CR leakage models do not explain this transition and induce an unavoidable jump in the mean free path of the CRs \citep{2012APh....39...12Z, 2009ApJ...694..951R, 1996APh.....5..367B}. \citet{2008AdSpR..42..486P} and \citet{2013ApJ...768...73M} propose a model to account for the escape of CRs from SNRs which is disconnected from the acceleration process. This model assumes that escaping CRs progressively leak out from the source which can be seen as a CR cloud, hence the name cosmic ray cloud (CRC) model. The model further considers that while escaping into the ISM, CRs propagate along the background magnetic field and that the cloud expansion is sufficiently slow to approximate the CR transport by a 1D process (see \citet{2008AdSpR..42..486P, 2013ApJ...768...73M}). Cosmic rays, whilst propagating along the background magnetic field, trigger slab-type Alfv\'en waves through the resonant streaming instability. The generation of self-generated turbulence produces a reduction of the CR diffusion coefficient. Finally, the model is developed in the quasi-linear approximation, so it is valid only in the case of well-developed, weak turbulence, an assumption that requires verification a posteriori. The 1D approximation for the CR transport is valid for distances to the CRC that are shorter than the background magnetic turbulence coherence length.
    
    In this paper, we use the kinetic code developed and presented in N16 and N19. We adapt the simulations to the case of partially ionized phases. We again use the setup proposed by \citet{2013ApJ...768...73M} and N16. We describe the transport of CRs and waves along magnetic field lines by solving two coupled equations given by
    
    \begin{equation}
    \label{eq:CRs}
        \pdv{P_{\rm CR}}{t} + V_{\rm A} \pdv{P_{\rm CR}}{z} = 
        \pdv{}{z}\left( D \pdv{P_{\rm CR}}{z} \right)
    ,\end{equation}
    
    \noi for the CR pressure $P_{\rm CR}$, and 
    
    \begin{equation}
    \label{eq:waves}
        \pdv{I}{t} + V_{\rm A} \pdv{I}{z} = 2 \left(\Gamma_{\rm growth}  - \Gamma_{\rm d}\right) I
        + Q 
    ,\end{equation}
    
    \noi for the waves energy density $I$. Equation (\ref{eq:CRs}) governs the evolution of the CR pressure $P_{\rm CR}$ in time and space. The space propagation is controlled by two processes: an advection with the scattering centers at a speed $V_{\rm A}$ and a random walk along the background field with a diffusion coefficient $D$. We note that in our case both $V_{\rm A}$ and $D$ are energy dependent. The Alfv\'en speed depends on the energy regime (see section \ref{S:DAM}): if ions and neutrals are coupled, the Alfv\'en speed is calculated using the total density, while in the decoupled regime only the ion density should be retained. The Alfv\'en speed can also be space-dependent in case the ISM is inhomogeneous. In that case, the above equations have to be modified \citep{1975MNRAS.172..557S, 2012ApJ...745...35Z}. It should be stressed that the above equations are restricted to the quasi-linear theory. Any modelling of the back-reaction of the self-generated turbulence over the background turbulence is therefore beyond the scope of this simple model and requires the use of an MHD code properly coupled to CR kinetics. We return to this point in sections \ref{S:GRA} and \ref{S:COC}. The diffusion coefficient is $D(E) = D_{\rm B}(E)/I(E)$ where $D_{\rm B}(E)$ is the Bohm diffusion coefficient (see \citet{1978MNRAS.182..147B, 1978MNRAS.182..443B}) and $I(E)$ is the energy density of the resonant Alfv\'en waves. The background diffusion coefficient is defined by $D_0 = D_{\rm ISM} (E / 10~\rm{GeV})^{0.5}$ and has been fixed to $D_{\rm ISM} = 10^{28}~\rm{cm}^2~\rm{s}^{-1}$ which is of the order of the value inferred from direct 
measurements of the ratios of CR primary to secondary. Cosmic ray transport is linked to the turbulence by equation (\ref{eq:waves}). In the latter, the growth of Alfv\'en waves is governed by the relaxation of the streaming instability which corresponds to the first RHS term. \citet{1975MNRAS.172..557S} gives the expression for the growth rate:
    
    \begin{equation}
        2 \Gamma_{\rm growth} I = - \frac{V_{\rm A}}{W_0} \pdv{P_{\rm CR}}{z}
    ,\end{equation}
    
    \noi where $W_0 = B_0^2/8 \pi$ corresponds to the magnetic energy of the medium. The second RHS term in equation \ref{eq:waves} describes the damping of the waves due to ion-neutral collisions and turbulent damping $\Gamma_{\rm d} = \Gamma_{\rm IN} + \Gamma_{\rm turb,L16}$. We also add the nonlinear Landau effect to handle cases where $I(E)$ becomes sufficiently close to one (see N16 and N19). Finally, the last term allows to consider all external sources of turbulence, and we set $Q = 2 \Gamma_{\rm d} I_0$ where $I_0$ is the background turbulence level. Hereafter, the latter is assumed to be identical whatever the ISM phase. A detailed investigation of the large-scale-injected turbulence in partially ionized media is a difficult task and is postponed to a future study (see the discussion in \citet{2016ApJ...826..166X}). 
    
    \noi The physical CR pressure is defined by 
    
    \begin{equation} 
        P_{\rm CR}  = \frac{4 \pi}{3 c^3} E^4 f(E)
    ,\end{equation}
    
    \noi where $f(E)$ is the CR energy space phase function defined by 
    
    \begin{equation}
        f(E) = \frac{3c^3}{16 \pi^2 R_{\rm s}^3} \frac{1}{E^2} \dv{N}{E}
    ,\end{equation}
    
    \noi where $R_{\rm s}$ is the shock radius and the CR energy spectrum $\dv*{N}{E}$ is given by
    
    \begin{equation}
        \dv{N}{E} = \frac{(2- \Gamma) W_\mathrm{CR}}{E_\mathrm{max}^{2-\Gamma} - E_\mathrm{min}^{2-\Gamma}} E^{-\Gamma}
    ,\end{equation}
    
    \noi if $\Gamma \neq 2$ and 
    
    \begin{equation}
        \dv{N}{E} = \frac{W_\mathrm{CR}}{\ln{(E_\mathrm{max})}- \ln{(E_\mathrm{min})}} E^{-\Gamma}
    ,\end{equation}
    
    \noi if $\Gamma = 2$. Here, $W_\mathrm{CR} = 10^{50}~\mathrm{erg}$ is the amount of energy of the SN explosion imparted into the CR pressure, and $E_\mathrm{min} = 2\times 10^{-4}~\mathrm{GeV}$ and $E_\mathrm{max} = 2 \times 10^{4} ~ \mathrm{GeV}$ are the energy limits of the CR spectrum. 
    
    Cosmic rays are initially confined in a region of size $a = 2 R_{\rm s}/3$ (see footnote 1  of N16 for an explanation) and an initial pressure $P_{\rm CR}^0$. Outside, the initial CR pressure is negligible. At the beginning of the simulation, the energy of turbulent waves is equal to its background value $I_0$. Following \citet{2013ApJ...768...73M}, we introduce the partial pressure $\Pi$ given by 
    
    \begin{equation}
        \Pi = \frac{V_{\rm A}}{D_{\rm B}} \Phi_{\rm CR}
    ,\end{equation}
    
    \noi where $\Phi_{\rm CR} = \int_0^\infty \dd{z} P_{\rm CR} = a P_{\rm CR}^0$ corresponds to the CR flux across a tube section. The physical meaning of $\Pi$ is discussed in detail in N16. We can identify two regimes of CR propagation depending on the value of $\Pi$: 
    \begin{itemize}
        \item If $\Pi > \rm{max}(1, \tau_{\rm diff}/ \tau_{\rm damp})$ where $\tau_{\rm diff} \approx a^2/D$ corresponds to the CR diffusion time-scale and $\tau_{\rm damp} \approx \left[2\Gamma_{\rm d} \right]^{-1}$ is the damping time scale of waves. Cosmic ray transport is nonlinear and the streaming instability is efficient enough to produce slab waves. 
    \item If $\Pi < \rm{max}(1, \tau_{\rm diff}/ \tau_{\rm damp})$. Cosmic ray transport is linear, the CR flux is too low to produce waves and the diffusion proceeds with $D = D_{\rm ISM}$.
    \end{itemize}
    
    If in that latter case in addition we neglect advection, the system of equations (\ref{eq:CRs}) and (\ref{eq:waves}) becomes 
    
    \begin{equation}
        \pdv{P_{\rm CR}}{t} = D_{\rm ISM} \pdv[2]{P_{\rm CR}}{z}
    ,\end{equation}
    
    \noi where $D_{\rm ISM} = D_{\rm B}/I_{\rm ISM}$ and 
    
    \begin{equation}
        \pdv{I}{t} = 0 
    ,\end{equation}
    
    \noi which means that $I = I_{\rm ISM}$. This situation is systematically achieved after a sufficiently long time as the flux of CRs across the tube section decreases and the condition $\Pi < \rm{max}(1, \tau_{\rm diff}/ \tau_{\rm damp})$ is realized. In this case, the analytic solution is given by 
    
    \begin{equation}
        \label{eq:TP}
        P_{\rm CR}(z, t) = \frac{1}{2} P_{\rm CR}^0 \left[ 
                              \rm{erf} \left( \frac{a -z}{2 \sqrt{D_{\rm ISM}t}} \right) +
                              \rm{erf} \left( \frac{a +z}{2 \sqrt{D_{\rm ISM}t}} \right)
        \right].
    \end{equation}
    
    \noi
   
   In the above framework, \citet{2013ApJ...768...73M} defined the CRC half-life time $t_\frac{1}{2}$ as the time it takes for the CR integrated pressure in the initial sphere to be divided by two. The CRC half time is defined by the relation 
    
    \begin{equation} 
        \label{eq:half-life time}
        \int_{-a}^{+a} P_{\rm CR}(z,t_\frac{1}{2}) \dd{z} = 2 a \frac{P_{\rm CR}^0}{2}
    ,\end{equation}
    
    \noi where $P_{\rm CR}^0$ is the initial CR pressure in the cloud. To each CR energy corresponds a particular half-life time. In order to determine at which moment and at which SNR radius a given population of CR escapes, N16 proposes to intersect the half-life time evolution function resulting from the equation (\ref{eq:half-life time}) $t_\frac{1}{2}(a)$ with the solution $t(R_{\rm SNR})$ describing the time evolution of the radius of an SNR. The approach adopted here is more relevant for type Ia SNe which evolve in an unperturbed ISM contrary to the case of core-collapse SNe where the wind of the stellar progenitor shapes the ambient medium. However, if the stellar wind magnetic field has a relatively homogeneous structure over scales that are larger than $R_{\rm SNR}$ one may still treat the escape problem as 1D but now using the path length along magnetic field lines. \\
    
    The evolution of SNRs has been described by \citet{1988ApJ...334..252C, 1999ApJS..120..299T} and can be decomposed into four expansion stages : 
    
    \begin{itemize}
        \item Free expansion,  
                \begin{equation*} t<t_{\rm free} \Rightarrow t \propto R_{\rm SNR}; \end{equation*} 
        \item Sedov-Taylor (ST),  
        \begin{equation*} t_{\rm free} < t < t_{\rm PDS} \Rightarrow t \propto R_{\rm SNR}^{5/2}; \end{equation*}
        \item pressure-driven snowplow (PDS),  
         \begin{equation*} t_{\rm PDS} < t < t_{\rm MCS} \Rightarrow t \propto R_{\rm SNR}^{10/3}; \end{equation*} 
        \item momentum-conserving snowplow (MCS),  
        \begin{equation*} t_{\rm MCS} < t < t_{\rm merge} \Rightarrow t \propto R_{\rm SNR}^4. \end{equation*} 
    \end{itemize}
    
\begin{figure*} \centering 
                
                                \includegraphics[width=1.0\textwidth]{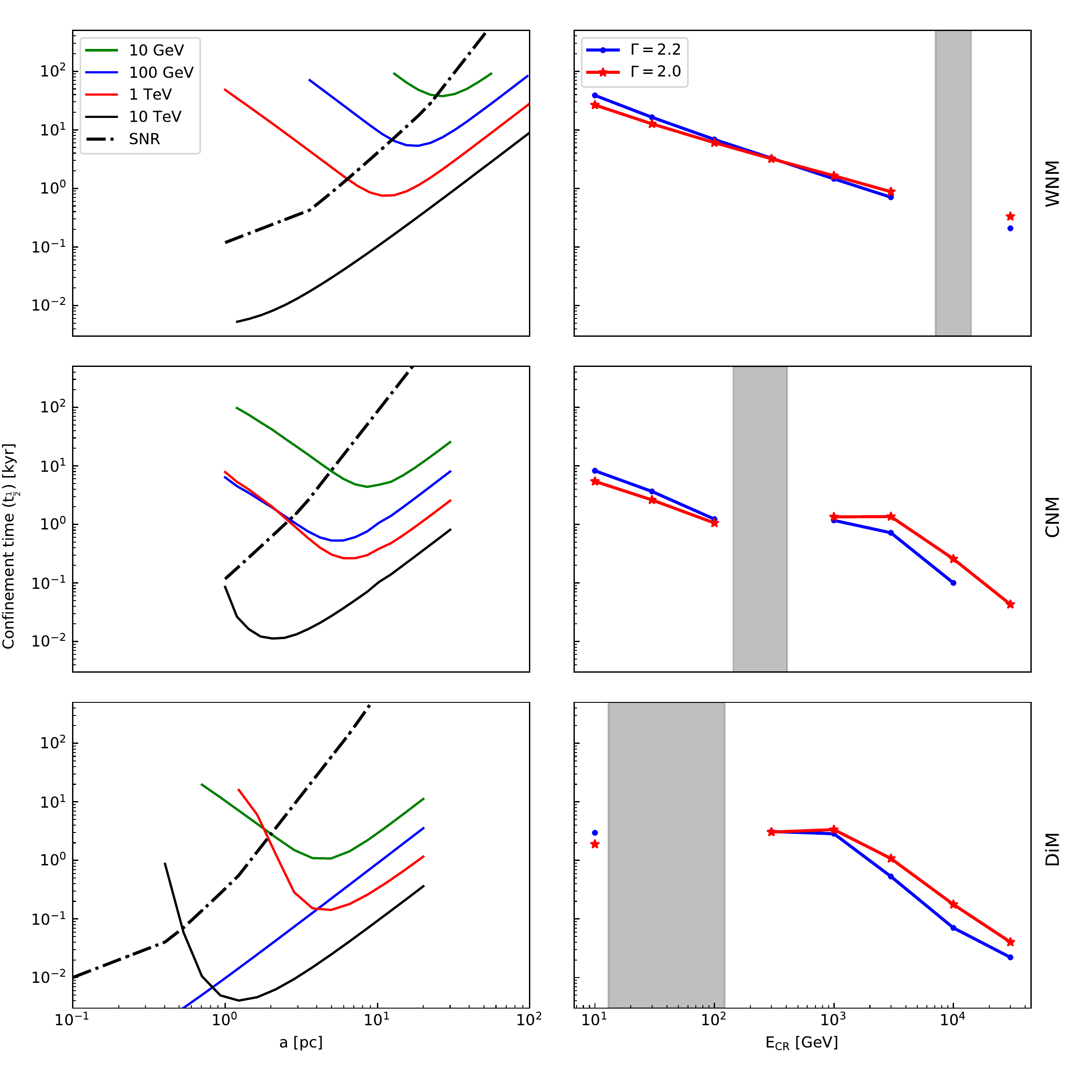}

                \caption{\label{fig:half-time} Properties of the expansion of the CRC in the ISM. \textit{Left}: Evolution of the half-life time of the CRC as a function of its initial radius $a$ in parsec units. Green, blue, red, and black solid lines refer to particle energies at $10$, $10^2$, $10^3$, and $10^4$ GeV. The black dashed line refers to the time evolution of the radius of the SNR front shock in the considered phase: WNM for the top figure, CNM for the middle one, and DiM for the bottom one. These plots have been obtained for $\Gamma = 2.2$. The right part of the figure represents the evolution of the half-life time of the CRC derived from the left plots as a function of CR energy. The blue solid line refers to a CR energy spectral index of $2.2$ while the red dashed line refers to a spectral index of $2.0$. The gray zone corresponds to the no-wave propagation zone; it does not lead to any $(a, t_{\rm 1/2})$ solution.}
            \end{figure*}

    \noi The constant of proportionality is defined by the radius of the SNR at the transition between the free and ST stages (see N16), at this stage the ambient medium starts to dominate the SNR dynamics.  
    
    \begin{equation}
    \label{eq:R_SNR_Nava}
        R_{\rm SNR} = 5.0 \left( \frac{E_{\rm{SNR},51}}{n_T} \right)^\frac{1}{5} \left[ 1 - \frac{0.05 M_{\rm{ej},\odot}^\frac{5}{6}}{E_{\rm{SNR},51}^\frac{1}{2} n^\frac{1}{3} t_{\rm kyr}} \right]^\frac{2}{5} t_{\rm kyr}^\frac{2}{5} ~ \rm{[pc]}.
    \end{equation}
    The SNR radius depends on the SN mechanical energy $E_{\rm SN}$ in units of $10^{51}$ ergs, the ejecta mass $M_{\rm ej, \odot}$ in units of solar mass, the ambient density $n_{\rm T}$ in units of $\rm{cm}^{-3}$, and the time in kiloyears.\\
    In order to understand the properties of the CR escaping process, we determined for each phase (WNM, CNM and DiM) and for each population of CRs (10 GeV, 100 GeV, 1 TeV, 10 TeV) the time at which half of the initial CR pressure has escaped from the CRC and to what radius the
shock extends. Our results are presented in figure \ref{fig:half-time}.

    \paragraph*{Warm neutral medium}: In the WNM phase, we recover the results obtained in N16 (see figure \ref{fig:half-time} upper row). We observe that low-energy CRs escape late, at 10 GeV $R_{\rm esc} \approx 25~\rm{pc}$ and $t_\frac{1}{2} \approx 42~\rm{kyr}$, at the start of the radiative stage. As energy increases, the radius and escape time decrease linearly. We note that the escaping process is nonlinear ($\Pi \approx 7\times 10^3, ~2 \times 10^3, ~0.5 \times 10^3 \gg 1$) at 10, 100, and 1 000 GeV respectively meaning that the self-generated turbulence can --a priori-- confine particles near the source. However, even if the growth rate is high it has to be compared with the damping rate, which can itself also be high. This comparison is the object of section \ref{S:RES}. We do not find any escaping solution for 10 TeV CRs. Figure \ref{fig:damping} shows that Alfv\'en waves do not propagate, hence particles cannot trigger any waves and their propagation is mostly controlled by perturbations at scales larger or equal to their Larmor radius (\citet{2004ApJ...614..757Y}).

    \paragraph*{Cold neutral medium}: In the CNM phase, the description of the CRC expansion is more complex (see middle
row of Fig. \ref{fig:half-time}).  Cosmic rays escape earlier, but the radiative stage of the SNR occurs also earlier. Cosmic rays at 10 GeV escape the latest ($\sim 9~\rm{kyr}$) at a shock radius of $\sim 5~\rm{pc}$. The solutions have $\Pi \approx 31 \times 10^3, ~ 4 \times 10^3, ~ 25, ~ 4 \gg 1$ or $>1$ at 10, 100, 10$^3$ and 10$^4$ GeV, respectively. At energies of 100 GeV and 1 TeV, CRs escape at a similar time ($\sim 1.2~\rm{kyr}$) and radius ($\sim 2.7~\rm{pc}$). This is due to the fact that the damping rate of 1 TeV waves occurs in the coupled ion-neutral collision regime and hence drops by a factor 100 with respect to the damping of low CR energy. At 10 TeV, CRs are released at a small radius ($<~2~\rm{pc}$), hence their pressure gradient is still high. 
    
    \paragraph*{Diffuse molecular medium}: In the DiM phase, the self-generated turbulence has larger amplitudes than in the CNM and the WNM (see figure \ref{fig:half-time}, lower row). This is essentially due to the small escaping CR radius which is a consequence of SNR dynamics in a medium  with a high density. Generally, CRs in the DiM escape earlier than in the CNM and they all escape during the radiative phase, which also occurs earlier. At 1 TeV, CRs seem to escape slightly before 10 GeV CRs. Similarly to the case of CNM, this is due to the difference between the damping rates in the coupled and decoupled regimes. Figure \ref{fig:damping} also shows that the turbulence in resonance with 100 GeV CRs is immediately damped explaining their linear diffusion. The half-life time of the CRC at 10 TeV is very small ($\sim 0.07~\rm{kyr}$) and its  radius is also quite small ($< 1~\rm{pc}$). This produces substantial turbulence; values of $\Pi$  are about $97 \times 10^3$, $16,$ and $9$ at 10, 10$^3$, and 10$^4$ GeV, respectively. \\
    
        It is also possible to study the CRC expansion properties by observing the cloud half-life time evolution as a function of the CR energy (see left parts of Fig. \ref{fig:half-time}). In the WNM, the half-life time decreases linearly with the energy of CRs. Low-energy CRs are efficiently confined while high-energy CRs propagate quasi linearly. In the colder phases (CNM, DiM), we observe a gap around the energies corresponding to the nonpropagating turbulence range followed by a plateau due to the substantial level of self-generated turbulence connected with the drop of the damping rate. At very high energies, the half-life time decreases again because CRs are injected with an energy spectrum with power-law of negative index, this means that the highest energies account for less pressure. We also performed simulations with a harder CR spectrum with $\Gamma = 2.0$ but keeping the same power imported into CRs, the half-life time decreases at low energy because of the decreasing value of the pressure but this effect is not strong (see the right-hand panels of Fig. \ref{fig:half-time}). At high energy the effect is inverted because of the generation of larger numbers of Alfv\'en waves due to the higher level of CR pressure. \\
    
   In Table (\ref{table:regime}) we summarize four different turbulence regimes operating in the phases under study. The production of self-generated turbulence can be seen to depend on the wave growth rate and the wave-damping rate imposed by the density and the ionization rate of the external medium. Let us briefly describe them.
    
    \begin{itemize}
        \item Regime 1 is that of strong nonlinear propagation. It represents the case where the wave growth rate is strong which is the case when the escaping radius is small, and the damping rate is reduced as in the coupled regime. This is the regime investigated by \citet{2013ApJ...768...73M}. The self-generated turbulence rate is growing almost exponentially with time leading to a strong CRs confinement. This regime is represented by a continuous decrease of the CR diffusion coefficient between $t_{1/2}$ and $4 t_{1/2}$ which is the typical limit time investigated in this study (see also N16 and N19).    
        \item Regimes 2 and 3 are transitory. Regime 2 corresponds to the case where the wave growth rate is high and the damping rate is strong, but still the difference between the two rates is small. Regime 3 corresponds to the case where the streaming instability is weak but produces waves weakly damped by interaction with the plasma. These regimes are represented by a decline followed by a growth of the CR diffusion coefficient between $t_{1/2}$ and $4 t_{1/2}$.
        
        \item Regime 4 is the regime dominated by the damping process. It represents the case where the waves growth rate is small which is the case when the CRC radius is large, and the damping rate is strong. This regime is represented by a continuous growth of the CR diffusion coefficient between $t_{1/2}$ and $4 t_{1/2}$.  
    \end{itemize}
    
        \begin{table} \centering 
    \label{table:regime}
                \begin{tabular}{c|cc}
                                              & $\frac{a^2I_0\Gamma_d}{D_B} \ll 1$ & $\frac{a^2I_0\Gamma_d}{D_B} \gg 1$ \\                            
        \hline
         $\frac{a^2I_0\Gamma_g}{D_B} \gg 1$   & $\dv{I}{t} \approx 2 \Gamma_g I$   & $\dv{I}{t} \approx 2(\Gamma_g  - \Gamma_d)I$ \\  
                                              & \rm{Regime} 1                 & \rm{Regime} 2 \\ 
        \hline
                 $\frac{a^2I_0\Gamma_g}{D_B} \ll 1$   & $\dv{I}{t} \approx 0$              & $\dv{I}{t} \approx - 2 \Gamma_d I$   \\  
                                              & \rm{Regime} 3                  & \rm{Regime} 4 
        \end{tabular}
        \caption{Table representing the different self-generated turbulence regimes as functions of the CR streaming instability growth rate and the Alfven waves damping rate by the medium. The expressions $a^2I_0\Gamma_d/D_B$ ($a^2I_0\Gamma_g/D_B$) represent the normalized damping (growth) rate. } 
    \end{table}

     \subsection{Alternative escape models}
In the model presented above and in N16 it is necessary to check that the shock speed is always larger than $110~\rm{km}~\rm{s}^{-1}$ which corresponds to the limit speed below which the shocked medium does not generate UV radiation and cannot ionize the precursor medium \citep{1979ApJ...227..131S}. In that case, ion-neutral collisions quickly damp any magnetic fluctuations 
(\cite{2017ApJS..229...34S}). We have checked in our simulations that the speed of the shock at the escape time of the lowest CR energies is always greater than this limit. \\
Another aspect associated with the ionization radiation produced by the shock wave is the possibility to have extended ionization fronts in the upstream medium \citep{2018AstL...44..769Z}. This effect is possibly expected for fast adiabatic shocks \citep{2008ApJS..178...20A}. Extended ionization precursors if they exist should rather be expected in the WNM, whereas in denser media their extension should be reduced by a factor 1/$n_{\rm H}$ where $n_{\rm H}$ is the hydrogen density. Balmer dominated shocks however show limited extension of a few percent of the SNR size of the heated gas precursors \citep{2007ApJ...659L.133L}.\\

However, the way CRs escape strongly depends on the evolution of the magnetic perturbations in the shock precursor, especially once the SNR has entered the radiative phase. We now consider two extreme cases of  CR release in the ISM which reflect our ignorance of the process. Hereafter, we refer to model $\mathcal P$ (for primary), the escape model presented in N16 and N19 and above. Model $\mathcal P$ is the model default hereafter unless otherwise specified. \\ 

In order to differentiate the escaping solutions obtained with the alternative models from the ones obtained with the main model, we refer to the SNR escaping radius variable as $a_{\rm esc}$ and the CR escaping time as $t_{\rm esc}$. These are analogous to the variables $a$ and $t_\frac{1}{2}$ but are obtained from different physical assumptions, as described below (they are identical in model $\mathcal P$).
     
                \paragraph*{Model $\mathcal F$  (fast)} : Magnetic field fluctuations are rapidly damped in the precursor once the shock enters the radiative phase implying that all accelerated particles are released at $a(E) = R_{\rm PDS}$ if $E < E^*$. Here, $E^*$ is the maximal energy of particles escaping at the start of the radiative phase in model $\mathcal P$, and depends on the properties of the external medium. As a consequence, we observe in Table \ref{table:alternative} that the radii and escaping times of CRs drop with respect to the values obtained in model $\cal P$. As escaping radii are smaller, the wave growth rate increases leading to stronger nonlinear propagation effects with respect to model $\mathcal P$.
        
        \paragraph*{Model $\mathcal S$ (slow)} : We suppose that the shock precursor is sufficiently ionized. Once in the radiative phase, the fluid is compressed and all particles that have not yet escaped at $t_{\rm PDS}$ stay confined until the shock velocity drops below $110~\rm{km}~\rm{s}^{-1}$. In this case we observe in Table \ref{table:alternative} that the radii and the escaping times of CRs are large and the propagation is less nonlinear with respect to the model $\mathcal P$.

     \begin{table} \centering 
     \label{table:alternative}
        \begin{tabular}{cccc|r}
                Phase                  &   WNM   &   CNM   &   DiM       &   \\
            $E_{\rm CR}$ [GeV]  &   $10$  &   $10$  &   $10-10^3$ &   \\ 
            \hline 
            $a_{\rm esc}$ (pc)               & $24.7$  & $4.95$  & $2.00-2.12$ & Model $\cal P$ \\ 
            $t_{\rm esc}$ (kyr)  & $42.1$  & $8.22$  & $2.99-3.46$ &         \\
            \hline
            $a_{\rm esc}$ (pc)               &  $20.8$ &  $3.14$ & $1.22$      & Model $\cal F$ \\ 
            $t_{\rm esc}$ (kyr)  & $24.2$  & $1.90$  & $0.510$     &         \\
            \hline 
            $a_{\rm esc}$ (pc)               & $28.0$  & $5.42$  & $2.33$      & Model $\cal S$ \\ 
            $t_{\rm esc}$ (kyr)  & $60.7$  & $11.0$  & $4.66$      &       
            
        \end{tabular}
        \caption{Here we show the values of the half-life time and their associated escaping radii for the three models in this study at low energy. We only retained CR energies where we have differences between the different models.}
     \end{table}
 
\section{Cosmic ray-propagation results}\label{S:RES}
We performed simulations of CR propagation in homogeneous realizations of WNM, CNM, and DiM phases at different CR energies: 10 GeV, 100 GeV, 1 TeV, and 10 TeV. We investigated the time-dependent transition between the CRC and the CR background. We note that in order to avoid any nonphysical pressure gradients at small escape times ($< t_\frac{1}{2}$) we smooth the initial pressure step by a hyperbolic tangent that is 1\% of the escape radius. We adopt absorbing boundaries at z=0 and $z=z_{\rm max} = 500$ pc. 
    
   The results are presented in the Appendix in figures (\ref{fig:pgd_WNM}-1, \ref{fig:pdg_WNM_2}-2, \ref{fig:pdg_WNM_3}-3) for the WNM, (\ref{fig:pdg_CNM}-1, \ref{fig:pdg_CNM_2}-2, \ref{fig:pdg_CNM_3}-3) for the CNM, and (\ref{fig:pdg_DiM}-1, \ref{fig:pdg_DiM_2}-2, \ref{fig:pdg_DiM_3}-3) for the DiM and for the three different escape models (model $\mathcal P$: 1, model $\mathcal F$: 2 and model $\mathcal S$: 3). The color-code refers to the different ISM phases: yellow for WNM, green for CNM, and orange for DiM. The figures are separated in three parts. The upper part represents the spatial distribution of the CR pressure. The black dotted line represents the initial CRC pressure, and the blue, green, and red lines represent the simulation state at $t_\frac{1}{2}/4$, $t_\frac{1}{2}$, and $4 t_\frac{1}{2}$ , respectively,  as in N16. We added when necessary two more solutions in brown and pink at $50 t_\frac{1}{2}$ and $10^3 t_\frac{1}{2}$ , respectively, at high energies in the CNM and DiM phases to account for the slow dilution of the CRC when the damping rate drops. Dotted lines represent test-particle (TP) solutions while full lines represent numerical solutions. The middle part represents the spatial evolution of the CR pressure gradient along the mean magnetic field at the three above times. The bottom part represents the spatial evolution of the diffusion coefficient with time compared to $D_{\rm ISM}$ drawn with a black dotted line. 
    
   In the following sections, for each setup, CR leakage properties are described and associated qualitatively to a simplified propagation model as described in Table \ref{table:regime}. The level of nonlinearity depends on the competition between the growth rate and damping rate. At $t_{1/2}$ we use "turbulence generation length" to refer to the width over which the turbulence growth rate exceeds the damping rate. This width is expected to decrease with time.

        \subsection{Warm neutral phase}
Figure \ref{fig:half-time} shows our solutions in the WNM phase for model $\mathcal P$. In this phase the CR escape begins at the start ST phase and ends at the start of the radiative phase. A general trend shows that the highest energies escape first, a result already obtained by N16. Figure \ref{fig:pgd_WNM} shows the evolution of the CR pressure, the CR pressure gradient, and the diffusion coefficient evolution in space at three different times, $t_{1/2}/4$, $t_{1/2}$, and $4t_{1/2}$.

\begin{itemize}
\item At 10 GeV, CRs escape late ($t_\frac{1}{2} \approx 42$ kyr, see figure \ref{fig:half-time}) when the shock radius is about 25 pc, at the beginning of the radiative phase. With the exception of early times, numerical solutions are close to TP solutions (see figure \ref{fig:pgd_WNM}). The maximum pressure gradient ($10^{-32}~\rm{erg}/\rm{cm}^4$) is reached at $t_\frac{1}{2}/4$ but is not strong enough to induce self-generated turbulence at rates much in excess of the background turbulence. This solution closely resembles regime 4, described above. The width of self-generated zone turbulence is about $\sim 50~\rm{pc}$.

\item At 100 GeV, CRs escape at $\sim 7$ kyr when the shock radius is about 13 pc in the ST phase. This implies that the CR pressure gradient is higher, especially at earlier times ($1.8$ kyr) where it reaches $\sim 10^{-30}~\rm{erg}/\rm{cm}^4$. At these early times, CRs generate a noticeable amount of turbulence but as the SNR shock expands, the CR gradient drops and our solutions converge to the TP case. Between $t_\frac{1}{2}$ and $4 t_\frac{1}{2}$ we observe that diffusion coefficients are close to the interstellar one. The width of generated turbulence is about $\sim 40~\rm{pc}$. Because the level of self-generated turbulence is at first high, it is best described by regime 2.
    
\item At 1 TeV, CRs escape at $\sim 1.5$ kyr when the shock radius is about 6.4 pc still during the ST phase. At early times, CRs have a high pressure gradient but the generated turbulence is low because the most important part of them escape later than $t_\frac{1}{2}$. At this time, the value of the CR pressure gradient is about $\sim 5 \times 10^{-31}~\rm{erg}/\rm{cm}^4$ leading to a diffusion coefficient lower than $10^{28}~\rm{cm}^2/\rm{s}$. However, this turbulence is quickly damped because ions and neutrals are in the weakly coupled regime. The width of generated turbulence is $\sim 40~\rm{pc}$. It is again best described by regime 2. 
    
\item In model $\mathcal F$, most  CRs escape before the beginning of the radiative phase. Only 10 GeV CRs escape at the beginning of this phase at $t_{\rm esc} \approx 24~\rm{kyr}$ with $a_{\rm esc} \approx 21~\rm{pc}$ similarly to model ${\cal P}$. As a consequence, the self-generated turbulence is stronger as showed in Fig (\ref{fig:pdg_WNM_2}). At $t_{\rm esc}/4,$ the diffusion coefficient is about $\sim 5~10^{26}~\rm{cm}^2/\rm{s}$ and increases gradually to $2\times 10^{27}~\rm{cm}^2/\rm{s}$ at $t_{\rm esc}$ to recover its interstellar value at $4 t_{\rm esc}$. The generated turbulence width is similar to model $\mathcal P$ but the ratio $\Gamma_g/\Gamma_d$ is larger. This case is best described by regime 2.
    
\item In the model $\mathcal S$, most CRs escape before the beginning of the radiative phase except CRs at 10 GeV which escape later during the transition $t_{\rm PDS}-t_{\rm MCS}$ in the radiative phase, that is, at $t_{\rm esc} \approx 60.7~\rm{kyr}$  when the shock radius is about $28~\rm{pc}$. Our results are shown in figure (\ref{fig:pdg_WNM_3}). In this case, the growth rate is weak and our solutions fall into regime 4. Numerical solutions follow thse of the  TP case. 
\end{itemize}  

In summary, we find results similar to those obtained in N16. We show that at 10 TeV, CRs propagate following the TP case. We find, as expected, no strong differences among the three escape models with the exception of the propagation of 10 GeV CRs where nonlinear effects are stronger in model $\mathcal F$ and weaker in model $\mathcal S$.

        \subsection{Cold neutral phase}
    
Figure \ref{fig:half-time} shows that in model $\mathcal P$, in the CNM, CRs from 10 GeV to 10 TeV escape between the end of the ST and the beginning of the radiative phase. The solutions $(a, t_{1/2})$ do not follow a linear trend as in the case of the WNM. Indeed, around 1 TeV, CRs are more strongly confined as can be seen in Fig. \ref{fig:pdg_CNM}. Compared to the WNM case, CRs escape earlier in the CNM because of the SNR shock dynamics (see equation \ref{eq:R_SNR_Nava}) as well as the CRC expansion properties whose essential parameters are the initial pressure and the value of the nonlinear diffusion coefficient. Test-particle solutions are identical for all phases (see equation \ref{eq:TP}).
\begin{itemize}
\item At 10 GeV, CRs escape at $t_{\rm 1/2} \approx 8~\rm{kyr}$ when the cloud size is small ($a = 5~\rm{pc}$). The SNR shock is in the radiative phase. Figure \ref{fig:pdg_CNM} shows that the solutions are relatively close to the TP ones. At $t_{1/2}/4,$ the CR pressure gradient is high and reaches $\sim 10^{-28}~\rm{erg}/\rm{cm}^4$ but then decreases rapidly to $10^{-30}~\rm{erg}/\rm{cm}^4$ at $t_{1/2}$ and the self-generated turbulence level drops accordingly. At $4t_{1/2}$, the self-generated Alfv\'en waves are completely damped and the linear propagation regime is recovered. This case is described by the regime 2.
    
\item At 100 GeV, CRs escape early ($t_{1/2} \approx 1.2~\rm{kyr}$) during the ST phase when the cloud size is about $2.7~\rm{pc}$. As shown in figure \ref{fig:pdg_CNM}, the CR pressure gradient is $\sim 10^{-28}~\rm{erg}/\rm{cm}^4$ and then slightly decreases until $t_{1/2}$ to finally drop below $10^{-30}~\rm{erg}/\rm{cm}^4$ at $4t_{1/2}$. Diffusion is substantially suppressed ($D \sim 10^{26}-10^{27}~\rm{cm}
^2/\rm{s}$) until $t_{1/2}$. After this time, the solution converges to the TP case. This case is described by regime 2. 

\item At 1 TeV, CRs escape at the same time as 100 GeV CRs ($\sim 1.15~\rm{kyr}$). The generated turbulence rate is the same as for 100 GeV CRs and the diffusion is highly suppressed ($D \sim 3 \times 10^{26}~\rm{cm}^2/\rm{s}$). The distance over which turbulence is generated however is different between the two energies ($\sim 40 ~ \rm{pc}$ at 1 TeV versus $\sim 20 ~\rm{pc}$ at 100 GeV). The associated turbulence regime is regime 1 at earlier times. Here the solutions are well approximated by the solutions of \citet{2013ApJ...768...73M}. At $4t_{1/2}$, waves are damped and the linear propagation regime is recovered.
    
\item At 10 TeV, CRs escape during the ST phase ($t_{1/2} = 0.1 ~ \rm{kyr}$) when the cloud size is about $1~\rm{pc}$. At $t_{1/2}/4,$ the CR pressure gradient is high, about $10^{-27}~\rm{erg}/\rm{cm}^4$, and the diffusion coefficient is $\sim 10^{27}~\rm{cm}^2/\rm{s}$. The pressure gradient then decreases gradually during the simulation while the diffusion coefficient decreases gradually from $\sim 10^{27}~\rm{cm}^2/\rm{s}$ to $10^{26}~\rm{cm}^2/\rm{s}$ at $50 t_{1/2}$. This effect is a consequence of the weak damping rate of waves at 10 TeV (see figure \ref{fig:damping}). At this energy, the damping time of waves is about 100 kyr. Suppressing the damping term in the equation is a good approximation to describe the propagation which is described by regime 1 and the solutions derived in \citet{2013ApJ...768...73M}. Furthermore, the distance at which significant turbulence is generated is large, about $\sim 100~\rm{pc}$. We find a ratio of $\delta B / B_0 \sim 0.8$. In order to prevent the level of self-generated turbulence from overtaking the quasi-linear regime we include the effect of nonlinear Landau damping and the effect of perpendicular diffusion by modulating the solution along the background magnetic field with a perpendicular dilution. At each time step, we calculate a new pressure given by
\begin{equation}
    P_{\rm CR}^{\rm new}(z,y=0,t)= {1 \over 2} P_{\rm CR}^{\rm old} \times \left(\rm{erf}(-\Delta y^2/4 D_{\perp} \Delta t)+ \rm{erf}(\Delta y^2/4 D_{\perp} \Delta t)\right) \ ,
\end{equation}
where $D_\perp = I^2 D_\parallel$ and the spatial step perpendicular to the mean magnetic field direction is $\Delta y = \Delta z = \sqrt{2C D_\parallel \Delta t}$, where $C=0.2$ is the CFL constant. In practice, the impact of the perpendicular diffusion is found to be limited. At late times, the CR distribution slowly converges to the TP solution. These solutions are however at the limit of validity of the quasi-linear theory. This point is discussed in section \ref{S:DQL}.

\item Figure \ref{fig:pdg_CNM_2} shows the solutions of model $\mathcal F$. Cosmic rays escape before the start of the radiative phase except at 10 GeV where the escape proceeds at the start of the radiative phase with $t_{\rm esc} \approx 1.9 ~\rm{kyr}$ when the shock radius is about $\sim 3~\rm{pc}$. In that case, numerical solutions are highly nonlinear over a large distance of $\sim 20~\rm{pc}$. Pressure gradients are high with ($(\pdv*{P_{\rm CR}}{z})_{\rm max} \approx 10^{27}~\rm{erg}/\rm{cm}^4$, and $D_{\rm min} \approx 5 \times 10^{24}~ \rm{cm}^2/\rm{s}$) testifying to a strong turbulence rate. In this model, the propagation regime is regime 2 but the growth rate exceeds the damping rate until $4 t_{\rm esc}$. 
    
\item Figure \ref{fig:pdg_CNM_3} shows the solution of model $\mathcal S$. Cosmic rays escape before the start of the radiative phase except at 10 GeV where CRs escape between the PDS and the MCS phase at $t_{\rm esc} \approx 11 ~\rm{kyr}$ when the cloud size is about $5.4~\rm{pc}$. Only the solution at $t_{\rm esc}/4$ shows nonlinear behavior and reduced diffusion coefficients ($\sim 10^{26}~\rm{cm}^2/\rm{s}$) over a spatial range of $20~\rm{pc}$. At later times the turbulence is gradually relaxed and we tend to the propagation in regime 2 where $\Gamma_d$ becomes higher than $\Gamma_g$. As expected however the solutions are less nonlinear with respect to model $\mathcal P$.
\end{itemize}

In summary, model $\mathcal P$ solutions at high energy (10 TeV) are in the nonlinear regime over a long time even if the growth rate decreases with time because the damping in the coupled regime has dropped. Comparing model $\mathcal P$ with the two other models we find that the results are the same for 100 GeV, 1  TeV, and 10 TeV CRs for the three escape models. However, at 10 GeV, CRs are strongly confined in model $\mathcal F$. At this energy, diffusion is strongly suppressed by a factor exceeding 1000 with respect to standard values. 

        \subsection{Diffuse molecular phase}

Figure \ref{fig:pdg_DiM} shows our results for model $\mathcal P$ in the DiM phase. We find that here, CRs escape slightly earlier, and  from a smaller cloud than in the CNM phase. This has important consequences on the level of self-generated turbulence and on the CRC expansion properties. The confinement time of CRs as a function of their energy does not evolve in a monotonic way. In particular we observe that at 1 TeV and 10 TeV CRs are well confined. 

\begin{itemize}
        \item At 10 GeV, CRs escape at $t_{1/2} = 3~\rm{kyr}$ during the radiative phase when the cloud size is about $2~\rm{pc}$. The solutions at $t_{1/2}$ and $4t_{1/2}$ deviate from the TP ones. Diffusion is suppressed ($D_{\rm min} \approx 5~10^{25} - 2~10^{26} ~ \rm{cm}^2/\rm{s}$) but the self-generated turbulence drops at $4 t_{1/2}$ and this case falls into regime 2 with $\Gamma_d > \Gamma_g$. The width over which turbulence is produced is about $17~\rm{pc}$.
        
    \item At 1 TeV, CRs escape relatively late ($t_{1/2} \approx 2.4 ~ \rm{kyr}$) during the radiative phase when the cloud size is about $\sim 1.9~\rm{pc}$. High CR pressure gradients suggest that self-generated turbulence levels are high ($\pdv*{P_{\rm CR}}{z}_{\rm max} \approx 10^{-28}~\rm{erg}/\rm{cm}^4$ at $t_{1/2}/4$ and $t_{1/2}$) which is confirmed by the values of the diffusion coefficients ($D_{\rm min} \approx 2~10^{26}~\rm{cm}^2/\rm{s}$ for $t_{1/2}/4$ and $t_{1/2}$). The turbulence level decreases very slowly at $4 t_{1/2}$ : $(\pdv*{P_{\rm CR}}{z})_{\rm max} \approx 10^{-29}~\rm{erg}/\rm{cm}^4$ and $D_{\rm min} \approx 10^{27}~\rm{cm}^2/\rm{s}$. This case is associated to regime 2 where $\Gamma_g > \Gamma_d$.  
    
    \item At 10 TeV, CRs escape very early ($t_{1/2} = 0.07~\rm{kyr}$) during the ST stage when the cloud size is about $0.75~\rm{pc}$. The growth rates are high with $(\pdv*{P_{\rm CR}}{z})_{\rm max} \approx 10^{-27}~\rm{erg}/\rm{cm}^4$ and $D_{\rm min} \approx 3 \times 10^{26} - \times 10^{26}~\rm{cm}^2/\rm{s}$. Turbulence production increases with time as in the case of 10 TeV CRs in the CNM and rapidly reaches a high turbulence regime with $\delta B/B \sim 0.8$. Nonlinear Landau damping and dilution by perpendicular diffusion are both included. This case corresponds to regime 1 where $\Gamma_g \gg \Gamma_d$. We note that the solution at $t=10^3 t_{1/2}$ shows a slight increases at $z \sim 400$ pc, this corresponds to a boundary effect which has no impact on the solution. 
    
    \item Figure (\ref{fig:pdg_DiM_2}) shows the solutions for model $\mathcal F$. Cosmic rays at 100 GeV and 10 TeV escape during the ST phase while those at 10 GeV and 1 TeV escape at the beginning of the radiative stage at $t_{\rm esc} \approx 0.5~\rm{kyr}$ when the cloud size is about $\sim 1.2~\rm{pc}$. Turbulence levels are higher than in the model $\mathcal P$: $(\pdv*{P_{\rm CR}}{z})_{\rm max} \approx 10^{-26}~\rm{erg}/\rm{cm}^4$ and $D_{\rm min} \sim 10^{25}~\rm{cm}^2/\rm{s}$ in the case of 10 GeV CRs. We observe that turbulence levels remain constant over a width of about $50~\rm{pc}$. These solutions correspond to propagation in regime 2 where $\Gamma_g > \Gamma_d$. Solutions at 1 TeV are strongly nonlinear. The maximum value of the CR pressure gradient is about $\sim 10^{-27}~\rm{erg}/\rm{cm}^4$ and remains constant throughout the simulation while the turbulence rate increases with time (at $t_{\rm esc}/4$, $D_{\rm min} \approx 2 \times 10^{25}~\rm{cm}^2/\rm{s}$ while at $4t_{\rm esc}$ this value becomes lower than $10^{25}~\rm{cm}^2/\rm{s}$) over a width of about $40~\rm{pc}$. The CRC then relaxes towards the TP solution at later times. The waves growth rate is highly dominant and is in line with the propagation regime 1.
    
    \item Figure \ref{fig:pdg_DiM_3} shows the solutions for model $\mathcal S$. At 100 GeV and 10 TeV, CRs escape during the ST phase while those at 10 GeV and 1 TeV escape during the radiative phase at $t_{\rm esc} = t_{\rm PDS} = 4.7~\rm{kyr}$ when the cloud size is about $2.3~\rm{pc}$. At 10 GeV, turbulence levels are relatively weak compared to model $\mathcal P$. Only CRs escaping before $t_{\rm esc}/4$ produce noticeable turbulence ($D_{\rm min} \sim 10^{26}~\rm{cm}^2/\rm{s}$). The damping rate dominates at later times implying that the propagation regime is regime 4. At 1 TeV, CRs produce turbulence until relatively late times but finally relaxe to the TP solution.  
\end{itemize}

In summary, we find the same trend as for the CNM at high energy: we believe the diffusion to be in the nonlinear regime because the damping rate drops. Comparisons between the models show that the results are the same for 100 GeV and 10 TeV CRs because they escape before the beginning of the SNR radiative stage and the shock velocity is higher than $150~\rm{km}/\rm{s}$. However, the  propagation properties of the 10 GeV and 1 TeV CRs depend on the escape model. In model $\mathcal F$ the turbulence level is higher and increases with time compared to the solutions of model $\mathcal P$. Model $\mathcal S$ shows less non-linear solutions, as expected. 

\subsection{Propagation at high energies}\label{S:DQL}
Here we comment over our solutions at 10 TeV in the CNM and DiM phases. At this energy diffusion is found to be highly nonlinear because of the drop of the damping term and because escape occurs with small CRC sizes. Relaxation towards the TP indeed occurs, although it  does take a longer time with respect to smaller energies. However, our solutions are at the limit of validity of the quasi-linear theory and should be considered with some care. These cases require that the 2D diffusion be properly taken into account. This effect should contribute to diluting the local CR gradient and to decreasing the level of self-generated turbulence. In the meantime we also note that CRs at this energy have a large diffusing flux, meaning that they can trigger a large current that can further amplify magnetic perturbations through the triggering of the nonresonant streaming instability \citep{Inoue19}, producing further confinement. These two aspects however require a much more demanding numerical effort. Finally, our injection model is crude and more specific modeling is required find the phase of the SNR evolution at which CRs at high energy escape. All these aspects will be investigated in a forthcoming study. 
\section{Discussion: Residence time and grammage calculations}\label{S:GRA}
The grammage of CRs is defined by the density column of matter crossed by CRs along their pathway in the ISM. This quantity is highly dependent on the particle transport properties. The grammage is defined by $X \approx 1.4 m_{\rm p} n_{\rm T} c \tau_{\rm res}$ where $\tau_{\rm res}$ is defined as the CR residence time. At galactic scales, this time is observationally inferred from secondary to primary ratios as is the case for boron (B) to carbon (C). However, this measurement is an average over the whole CR journey in the Galaxy and it includes the transit in the galactic disc and in the halo. In order to more appropriately calculate the CR pathway, it is important to evaluate the contribution of the propagation close to the CR sources where it is anticipated that the CR self-generated turbulence may control CR transport \citep{2016PhRvD..94h3003D}.

We consider two ISM configurations. First, we calculate the residence time in the case of homogeneous phases as a function of the particle energy in section \ref{S:MONO}, and subsequently we consider cases with multiphase ISM configurations in section \ref{S:MULT}.

\subsection{Single-phase studies}\label{S:MONO}
In our semi-cylindric 1D problem configuration the residence time is defined as the characteristic time for a CR to escape from the CRC to a fixed distance $z_*$. We derive $\tau_{\rm res}$ using a method based on the fact that the CR trajectory outside the CRC is dominated by diffusion (an assessment which we have verified). The residence time is calculated from an average of the square of the diffusion distance (see N19):

    \begin{equation}\label{EQ:Z*}
        z_*^2 = \frac{ \int_0^\infty z^2 P_{\rm CR}(E, z, \tau_{\rm res}) \dd{z}}
               {\int_0^\infty P_{\rm CR}(E, z, \tau_{\rm res}) \dd{z}}.
    \end{equation}

\noi Once $z_*$ is fixed, we deduce $\tau_{\rm res}$ by balancing the two terms of Eq.\ref{EQ:Z*}.  \\ 

\begin{figure*}
    \centering
    \includegraphics[width=0.5\linewidth]{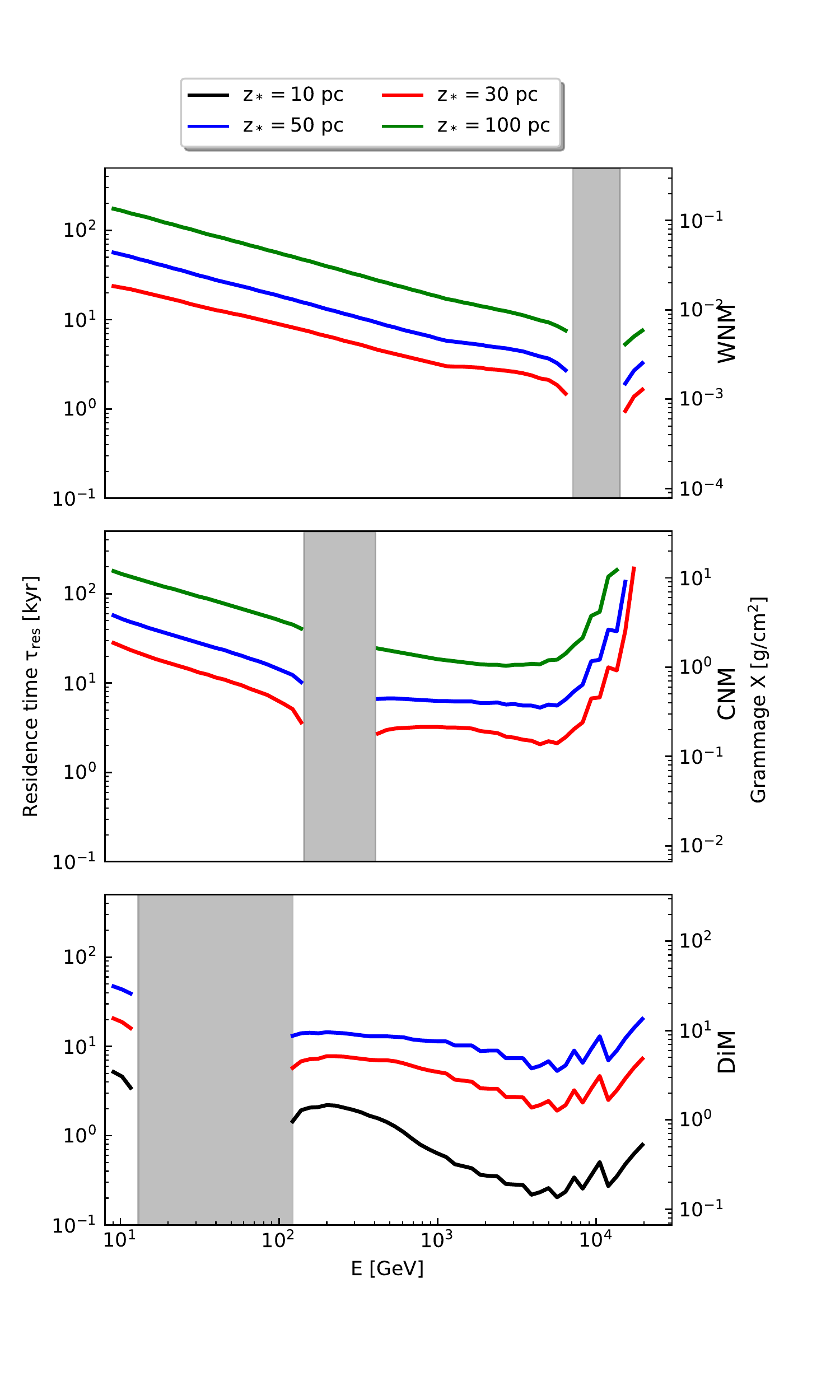}
    \caption{Grammage (right vertical axis) and residence time (left vertical axis) of CRs as a function of their energy in three different phases of the ISM : WNM (top), CNM (middle), and DiM (bottom). Solid lines show the results obtained at different distances $z_*$ from the center of the source: green (100 pc), blue (50 pc), red (30 pc), and black (10 pc). The shaded region marks the no-wave propagation zone. Oscillations at high energy in the CNM and DiM are due to the limited energy resolution used to reconstruct the different curves.}
    \label{fig:grammage_mono}
\end{figure*}

We calculate the CR residence time ($\tau_{\rm res}$) and the associated grammage ($X$) in a given ISM configuration. The results are presented in figure (\ref{fig:grammage_mono}) for the WNM (top), CNM (middle), and DiM (bottom). Solutions derived are represented by a continuous line. The positions $z_*$ from the cloud center are indicated by different colors: black ($10$ pc), red ($30$ pc), blue ($50$ pc), and green ($100$ pc). All calculations are realized in the framework of model $\cal{P}$ (see figure (\ref{fig:half-time})). 

\subsubsection{Warm neutral medium}
In the WNM at $z_* = 100 ~ \rm{pc}$ ($z_* = 50 ~ \rm{pc}$ and $z_* = 30 ~ \rm{pc}$ respectively), the grammage evolves as $E^{-0.5}$ and is about $X = 0.1 ~ \rm{g}/\rm{cm}^2$ ($X = 4 \times 10^{-2} ~ \rm{g}/\rm{cm}^2$ and $X = 2 \times 10^{-2} ~ \rm{g}/\rm{cm}^2$ , respectively) at 10 GeV and decreases linearly to reach $X = 10^{-2} ~ \rm{g}/\rm{cm}^2$ ($X = 2 \times 10^{-3} ~ \rm{g}/\rm{cm}^2$ and $X = 1 \times 10^{-3} ~ \rm{g}/\rm{cm}^2$ respectively) at 3 TeV. \\
In the WNM, we find in section \ref{S:RES} (see also figure \ref{fig:pgd_WNM}) that nonlinear effects are weak at almost all energies because ion-neutral interactions proceed in the decoupled regime. This reflects the solutions obtained by the two methods as both show a trend with a grammage which can be parametrized as $X \simeq X_{\rm{WNM}} (E/10 \rm{GeV})^{-0.5}$, with $X_{\rm{WNM}} \simeq 0.1~\rm{g/cm^2}$. This grammage evolution is characteristic of the transport in the background turbulence.

\subsubsection{Cold neutral medium}
In the CNM at $z_* = 100 ~ \rm{pc}$ ($z_* = 50 ~ \rm{pc}$ and $z_* = 30 ~ \rm{pc,}$ respectively), the grammage evolves as $E^{-0.5}$ and is about $X = 10 ~ \rm{g}/\rm{cm}^2$ ($X = 3 ~ \rm{g}/\rm{cm}^2$ and $X = 1.5 ~ \rm{g}/\rm{cm}^2$ respectively) at 10 GeV. The grammage in this model is not defined around 300 GeV due to the Alfv\'en waves nonpropagation band (see figure \ref{fig:damping}). We then see a slight softening of the slope around 3 TeV and a hardening up to 10 TeV where the grammage reaches $X = 4 ~ \rm{g}/\rm{cm}^2$ ($X = 2 ~ \rm{g}/\rm{cm}^2$ and $X = 1 ~ \rm{g}/\rm{cm}^2$,  respectively).   

At low energy ($E < 300$ GeV), the grammage approximately follows $X \propto E^{-0.5}$ which is characteristic of the propagation in background turbulence. The discussion in section \ref{S:RES} and the results in figure \ref{fig:pdg_CNM} show a trend very similar in the CNM and WNM at these energies. We do not have any solution around 300 GeV associated to the forbidden propagation zone. In that energy regime, the transport is completely controlled by the background turbulence. Above 300 GeV we can explain the change of behavior by the fact that diffusion is suppressed (D$\sim 5 \times 10^{26}~\rm{cm^2}/\rm{s}$) until we reach the half-life time of the CRC. At high energies, the diffusion is even more suppressed over larger scales and longer times which results in an increase of the grammage. We emphasize however that the turbulence levels obtained in this configuration are at the limit the quasi-linear CR transport theory, and more realistic calculations (including 2D effects, improved escape model, and the triggering of other types of instabilities) should lead to a more accurate estimation.

\subsubsection{Diffuse molecular}
In the DiM at $z_*= 50 ~ \rm{pc}$ ($z_* = 30 ~ \rm{pc}$ and $z_* = 10 ~ \rm{pc,}$ respectively), the grammage value is about $X = 40 ~ \rm{g}/\rm{cm}^2$ ($X = 20 ~ \rm{g}/\rm{cm}^2$ and $X = 4 ~ \rm{g}/\rm{cm}^2,$ respectively) at 10 GeV. We do not have any values of the CR grammage between 20 and 100 GeV because of the absence of escaping solutions due to the nonpropagation of Alfv\'en waves. Beyond 3 GeV, we observe a softening up to 10 TeV for $z_* = 30 ~ \rm{pc}$ and  $z_* = 50 ~ \rm{pc}$. Close to the source ($z_* = 10 ~ \rm{pc}$) we observe that the grammage decreases more strongly up to 3 TeV and then hardens between 3 and 10 TeV, where it finally reaches $X = 5 ~ \rm{g}/\rm{cm}^2$ ($X = 1.5 ~ \rm{g}/\rm{cm}^2$ and $X = 0.2 ~ \rm{g}/\rm{cm}^2,$ respectively).

In the DiM, we observe that the grammage evolution with the CR energy is modified compared to the low-energy CNM values and those in the WNM. Softer slopes are induced by stronger CR confinement at high energy (see \ref{fig:pdg_DiM}). The confinement effect is also visible in the CR pressure distribution.  

\subsubsection{Single-phase study}
In the WNM, the grammage does not exceed $0.1 \rm{g/cm^2}$ at 10 GeV and scales as $E^{-0.5}$. Still at 10 GeV, in the CNM and DiM the grammage is higher, namely $\sim 10~\rm{g/cm^2}$ for the CNM and $\sim 20~\rm{g/cm^2}$ for the DiM. These values are large, but it should be borne in mind that the CNM and DiM cover only a small fraction of the galactic disk volume, hence these media likely only contribute to a small fraction of the total grammage of the CR detected on Earth. Nonetheless, these results also show that the CR grammage may vary considerably depending on the location in the ISM. This could have implications for the production of light elements in the cold ISM located close to a CR source. The unexpected result we find is obtained at high energy where the grammage increases with the energy in the CNM and DiM. High values are found at 10 TeV of the order of $2~\rm{g/cm^2}$ and $5~\rm{g/cm^2}$ at 50 pc from the source in the CNM and DiM, respectively. Here again, due to the small volume-filling factor of these phases, we do not expect to have a strong contribution to the CR grammage at these energies. Furthermore, these values are likely overestimates since the results at these energies and in these phases are at the limit of the validity of quasi-linear theory. Finally, the results are obtained for a homogeneous ISM. More realistic multi-phase calculations are presented below for low CR energies and in a forthcoming work for all CR energies.  

\subsection{Hadronic collision losses}
While propagating in dense media, CRs are subject to proton-proton (pp) collisions and energy losses. The typical loss time by pp interaction is 
\begin{equation}
t_{\rm loss,pp}=\left(K_{\rm pp} \sigma_{\rm pp} n_{\rm H} c \right)^{-1} \ ,   
\end{equation}
where $K_{\rm pp} \simeq 0.2$ and $\sigma_{\rm pp} \sim 30$ mbarn is the pp collision cross section at the interaction threshold and is only increasing with the logarithm of the particle kinetic energy \citep{Kafexhiu14}. We have checked that even in the DiM at high energy we always find $t_{\rm res}/t_{\rm loss, pp} \ll 1$. Cosmic rays then propagate in dense phases without suffering strong losses, which can then be a fortiori neglected in Eq.\ref{eq:CRs}.

\subsection{Multiple-phase studies}\label{S:MULT}
We extend our study to more realistic ISM environments by calculating the CR grammage in an inhomogeneous ISM. We restrict our analysis to CR energies at 10 GeV for three reasons: 1) it is in that energy domain that the grammage has been primarily derived from secondary to primary direct measurements, 2) in this energy domain the self-generated turbulence possibly controls the CR propagation in the ISM \citep{2012PhRvL.109f1101B}, and therefore this is a good regime to test the impact of such turbulence over CR propagation in different media, and 3) in the decoupled ion-neutral regime the Alfv\'en speed is to be taken with respect to the ion density; it does not vary significantly from one phase to another. This means that CRs do not suffer from strong adiabatic losses while propagating from one medium to another, and therefore Eqs. \ref{eq:CRs} and \ref{eq:waves} are still a good approximation to describe CR and wave-coupled evolution. At high energies, ion-neutral collisions occur in the coupled regime, and the Alfv\'en speed is to be taken with respect to the total gas density. The Alfv\'en speed varies considerably and then the kinetic equations for particles and waves have to be generalized. This requires a more complete modeling, which is postponed to a future work.\\

For the multiple-phase case, we propose two setups from the least-dense to the densest phase: 

\begin{itemize}
    \item Setup 1 : The CRC evolves first in a WNM phase up to a distance of 50 pc where it reaches a CNM phase up to 130 pc. Beyond this distance, the WNM is retrieved up to 200 pc which is the limit of our simulation box. 
    \item Setup 2 : This setup is structured in the same way as Setup 1 except that we set a DiM cloud in the middle of the CNM phase, between 80 and 100 pc.
\end{itemize}

\begin{figure*}
    \centering
    \includegraphics[width=1.0\linewidth]{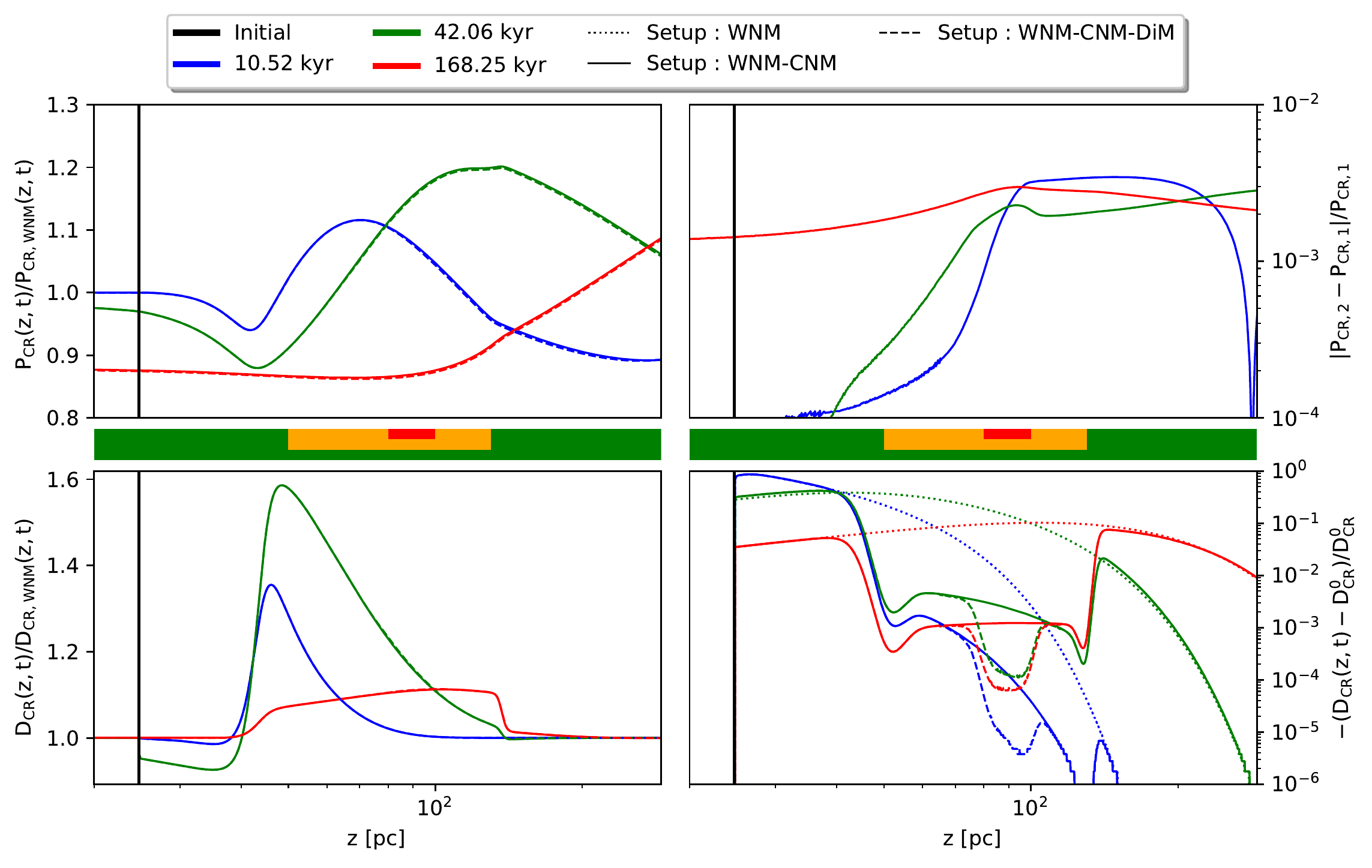}
    \caption{Evolution of the CR pressure (top figures) and their associated diffusion coefficient (bottom panels) in space at different times of the simulation based on the CRC half-life time for 10 GeV CRs in the WNM: 10.52 kyr (blue), 42.06 kyr (green), and 168.25 kyr (red). The simulation setups are represented by the different line styles: solid line for the WNM-CNM, dashed line for the WNM-CNM-DiM, and dotted line for the homogeneous WNM. The vertical black lines represent the initial size of the CRC. The phase decomposition for each setup is represented between the figures by the colored lines: green for WNM, orange for CNM, and red for DiM. The top-left figure shows the relative evolution of the CR pressure for WNM-CNM and WNM-CNM-DiM setups compared to the CR pressure evolution in the case of propagation in a homogeneous WNM phase. The top-right panel shows the relative CR pressure evolution for the WNM-CNM-DiM ($P_{\rm CR,2}$) and the WNM-CNM ($P_{\rm CR,1}$) setups, respectively. The bottom-left panel shows the relative evolution of the CR diffusion coefficients compared to the evolution of the CR diffusion coefficient in the case of propagation in a homogeneous WNM phase. The bottom-right panel shows the normalized relative evolution of the CR diffusion coefficient around the CNM and CNM-DiM phases. We recall that $D_{\rm CR}^0 = 10^{28}~\rm{cm}^2/\rm{s}$.} 
    
    \label{fig:multiphase}
\end{figure*}
\begin{figure*}
    \centering
    \includegraphics[width=1.0\linewidth]{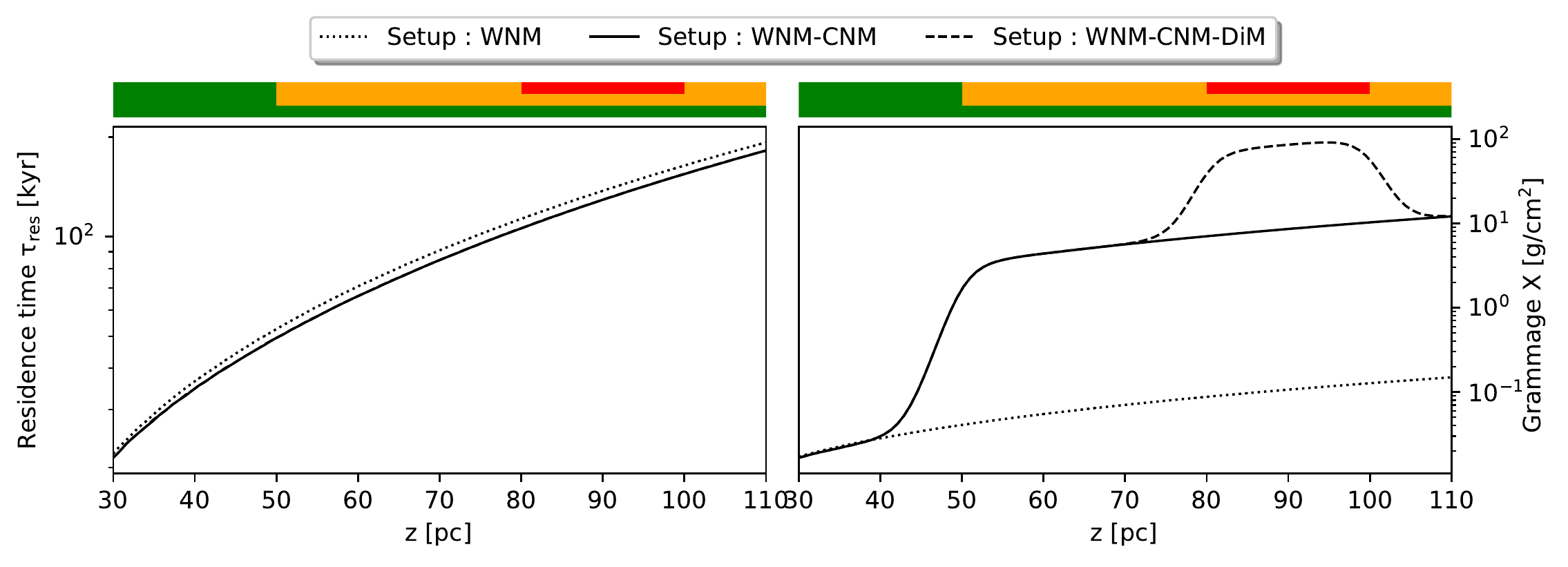}

    \caption{Evolution of the CR grammage (right part) and residence time (left part) in Setup 1: WNM-CNM and Setup 2 : WNM-CNM-DiM represented by a solid curve and a dashed curve, respectively. These setups are compared with the case of the expansion of a CRC in a homogeneous WNM represented by the dotted lines. Each setup phase configuration is represented by a color code: WNM (green), CNM (orange), and DiM (red).}

    \label{fig:grammage_multi}
\end{figure*}

\noi In order to avoid strong pressure and diffusion coefficient discontinuities between phases, all space-dependent variables appearing in Eqs (\ref{eq:CRs}-\ref{eq:waves}) are smoothed with a hyperbolic tangent function of a width equal to 3 pc. This transition can be changed and we verified that it does not affect the overall results of the grammage calculation. Our results are presented in figures \ref{fig:multiphase} and \ref{fig:grammage_multi} for the space-time evolution of the CR pressure and diffusion coefficient and for the grammage, respectively. In both figures, for each setup, WNM, CNM, and DiM phases are represented by green, orange, and red colors, respectively. In figure \ref{fig:multiphase}, the plot in the top left hand corner shows the evolution of the CR pressure at times corresponding to those chosen for the simulation of the expansion of the CRC in the WNM at 10 GeV. The plot in the bottom left corner shows the evolution of the diffusion coefficient. Cosmic ray pressures and diffusion coefficients are normalized to the solution obtained in the WNM case. In the top right corner, CR pressures for Setup 1 (solid line) are compared to those of Setup 2 (dashed line). In the bottom right, the diffusion coefficients for the three setups (WNM: dotted line, WNM-CNM: solid line, WNM-CNM-DiM: dashed line) are compared to those of the TP solution. In figure \ref{fig:grammage_multi} the CR residence time is represented in the left plot while the CR grammage is shown in the right one.\\

In order to simplify our analysis we use $\alpha_i = P_{\rm CR,i}(z,t) / P_{\rm CR,WNM}(z,t)$ and $\beta_i = D_{\rm CR,i}(z,t) / D_{\rm CR,WNM}(z,t)$ where $i = 1,~2$: the ratio between CR pressure and diffusion coefficients of the setup $i$ with the CR pressure and diffusion coefficients in the case of a propagation in an homogeneous WNM (see figure \ref{fig:pgd_WNM}).

\subsubsection{Setup 1 : WNM-CNM} 

For this setup, Fig. \ref{fig:multiphase} shows that the CR pressure distribution is modified by $\sim$ 10\% compared to the case of propagation in a homogeneous WNM. At 10.52 kyr, the curve (blue solid line) follows the same behavior as in the homogeneous WNM case up to a distance of $\sim 30$ pc. Beyond $30$ pc the ratio $\alpha_1$ decreases and increases suddenly at the level of the transition WNM-CNM (50 pc). Here, $\alpha_1$ reaches $\sim 1.15$ around 70 pc before decreasing again and tends to $\sim 0.9$ after the transition CNM-WNM. At 42.06 kyr, we observe that $\alpha_1$ has a slightly lower value than 1 at the spatial origin of the simulation. The dip close to the WNM--CNM  transition is then accentuated ($\alpha_1 \sim 0.9$) and $\alpha_1$ increases again reaching its maximum ($\sim 1.2$) around 110-120 pc and finally decreases. At 168.25 kyr, $\alpha_1$ is lower than 0.9 and is almost flat except at the CNM--WNM transition where it begins to increase.

The diffusion coefficient evolution is also affected by the structure of the ISM. At 10.52 kyr, the ratio $\beta_1$ is constant (and equals 1) up to the WNM--CNM transition (50 pc) and starts to increase to reach $\sim 1.3$ before gradually decreasing to retrieve the value of 1 at the CNM--WNM  transition. At 42.06 kyr, the ratio $\beta_1$ is lower than 1 ($\sim 0.9$) and starts to increase up to the WNM--CNM transition to reach $\sim 1.6$ and decreases progressively to reach 1 near the CNM--WNM transition. At 168.25 kyr, $\beta_1$ is again close to 1 and starts to increase slowly from the WNM--CNM
transition to reach 1.1 close to the CNM--WNM  transition where it decreases abruptly to the value of 1 beyond. \\ 

The different behaviors between the case of a propagation in a homogeneous WNM and Setup 1 essentially come from the Alfv\'en waves damping timescales (see figure \ref{fig:damping}). In the WNM, the damping time is about 10 yr while in the CNM it is about 1 yr. In the master equations (\ref{eq:CRs}) and (\ref{eq:waves}), the right-hand term, $2(\Gamma_\mathrm{growth} - \Gamma_\mathrm{d})I,$ decreases from the WNM to the CNM. As $\Pi$ is a conserved quantity regardless of the phase, the figure can be explained easily: the ratio of the pressures starts at 1 in the WNM, then as the damping is more severe in the CNM, less waves are produced at the  WNM--CNM  transition. The CR pressure in the CNM goes first below and above the WNM solution because CRs diffuse faster in the CNM. Finally, the CNM solution catches the WNM ones beyond, but the ratio reaches 0.9 instead of 1 because of the conservation of $\Pi$. The peak in CR pressure in the CNM is slowly advected out with time as the solution relaxes to the TP case. The bottom right part of figure \ref{fig:multiphase} shows that the difference between the self-generated and background diffusion coefficients is not large, unlike in homogeneous cases. This small difference at 10 GeV is due to the strong damping effect of ion-neutral collisions.


The CR residence time and grammage are presented by solid lines in figure (\ref{fig:grammage_multi}) on the left and on the right, respectively. We observe that the residence time increases linearly from 22 kyrs at 30 pc to 200 kyr at 110 pc. The evolution curve of Setup 1 is slightly lower than that obtained in the case of a CR propagation in a homogeneous WNM. The associated grammage is relatively weak close to the source ($X \sim 0.02 ~\rm{g}/\rm{cm}^2$ at 30 pc) and increases linearly by following the residence time trend up to the WNM--CNM transition where it increases by a factor of $100 \sim n_{T,\rm{CNM}}/n_{T,\rm{WNM}}$ to reach the order of $10~\rm{g}/\rm{cm}^2$ at 110 pc from the source. \\ 


\subsubsection{Setup 2 : WNM-CNM-DiM} 

As presented by the left plots of figure (\ref{fig:multiphase}), the observed trends in Setup 2 are almost equivalent to those observed in Setup 1. The modifications associated to the presence of the DiM, although small, are visible in the right-hand plots. The CR transport properties are essentially modified by the presence of the CNM. The effect of the DiM is negligible ($\sim 0.2$\% of relative CR pressure variation). The only impact can be seen in figure (\ref{fig:grammage_multi}) where the grammage increases due to the density contrast with respect to Setup 1.

It is important to note again that we did not take into account adiabatic losses associated to the spatial dependence of the Alfv\'en velocity. 
This aspect is left out for a future, more detailed work which may reveal a noticeable effect of the presence of the DiM phase on the CR propagation properties in the ISM. 

\section{Discussion and conclusion} \label{S:COC}

In this study, we extend the work initiated in \citet{2016MNRAS.461.3552N} aimed at investigating the propagation of CRs in the ambient ISM surrounding SNRs. Here, we focus on cold, partially ionized atomic and molecular phases. To this aim, using the CRC model we simultaneously solve two 1D transport equations in 1D: an equation over the CR pressure $P_{\rm CR}$ (see eq. \ref{eq:CRs}) and an equation over the Alfv\'en slab waves energy density $I(k)$ (see eq. \ref{eq:waves}). We restrict our analysis to  a resonant interaction where the particle Larmor radius and the wave number verify $r_{\rm l} = k^{-1}$ in the quasi-linear approximation of CR transport; hence the lack of  strong feed-back over the background turbulence. We determined the characteristic confinement time of CRs depending on their energy and the ISM phases: WNM, CNM, DiM. We discuss three different escape models once the SNR enters in the radiative phase, either considering a continuous escape (model $\mathcal{P}$), an escape at the end of the free expansion phase (model $\mathcal{F}$), or an escape in the late radiative phase once the SNR forward shocks reach a speed of $\sim 150$ km/s (model $\mathcal{S}$). We performed CRC evolution simulations for monoenergetic CR populations from 10 GeV to 10 TeV in both ISM phases in order to probe the evolution of the CR pressure distribution $P_{\rm CR}(E,z,t)$; the CR pressure gradient distribution $\pdv*{P_{\rm CR}(E,z,t)}{z,}$ which is analogous to the force exerted by CR fluid on the plasma; and the CR diffusion coefficient $D(E, z, t),$ which is intimately linked to the rate of excited Alfv\'en modes. Using the numerical solutions obtained we derived the grammage around a single cloud either in a homogeneous ISM or an ISM composed of multiple phases.          

\subsection{Results}

\noi Our results show that the Alfv\'en turbulence generated by the streaming of CRs can have an important effect on CR acceleration and propagation over distances of between a few tens and one hundred parsecs. 

\begin{itemize}

\item[(1)] The CR escape solutions are presented in figure (\ref{fig:half-time}) for the three phases under study. We find a similar trend for the WNM as already discussed by N16 for the warm ionized phase. We show that in the CNM, at 1 TeV and 3 TeV, CRs escape at the same time as 100 GeV at $\sim 1$ kyr while at 300 GeV CRs escape earlier at $\sim 0.4$ kyr at the same time as 10 and 30 TeV. In the DiM, the radiative phase begins at $\sim 0.3$ kyr and at 10 GeV, 300 GeV, and 1 TeV, CRs escape during the radiative stage at $\sim 2$ kyr, while 30 TeV CRs escape at $\sim 0.05$ kyr. The larger escape times at high energy are explained by the drop of the ion-neutral collisions in the coupled interaction regime. 
\item[(2)] We find that a non-negligible quantity of Alfv\'en modes are generated in the CNM ($D \sim 5 \times 10^{27}~\rm{cm}^2/\rm{s}$ at 10 TeV) and in the DiM ($D \approx \times 10^{26}$ at 10 TeV) after the CRC half-life time. These modes contribute to an enhanced CR confinement with respect to the transport by background turbulence. The model $\mathcal F$ systematically shows stronger turbulence generation and longer confinement times with respect to the two others models.

\item[(3)] In the WNM we show that the grammage behavior follows the observationally measured behavior and can be parametrized as $X = X_{\rm WNM}(z,10 \rm{GeV}<E_0<10\rm{TeV})(E/E_0)^{-0.5}$. The findings concerning propagation regimes in the CNM and the DiM imply that the CR grammage around SNRs can be enhanced at very high energy in these phases. In the CNM, the grammage has two regimes: the first one being between 10 GeV and 3 TeV with $X = X_{\rm CNM}(z, 10\rm{GeV}<E_0<3\rm{TeV})(E/E_0)^{-0.5}$. Above 3 TeV, the grammage increases. In the DiM, the grammage decreases more slowly than in the WNM from 300 GeV to 10 TeV. 

\item[(4)] We perform a study at 10 GeV with multiphase setups: WNM-CNM and WNM-CNM-DiM. We find that the CR grammage ($\tau_{\rm res}$) is controlled by the densest phase and can reach $\sim 10~\rm{g/cm^2}$ in the WNM-CNM case and up to $\sim 10^2~\rm{g/cm^2}$ in the WNM-CNM-DiM case. The CR pressure evolution is however controlled by the CNM phase. 

\item[(5)] As stated above, our solutions at high energy are at the limit of validity of the quasi-linear theory. A more complete modeling including a 2D description is necessary for more accurate estimations of the grammage. We can also argue that if the grammage is expected to be smaller as extrapolated from [B/C] measurements, it reflects an average of the ensemble of CR pathways in the ISM. Partially ionized phases like the CNM and the DiM (and denser phases) do not cover a large fraction of the galactic volume, meaning that in total they do not significantly contribute to the average grammage.

\end{itemize}

\subsection{Discussion and perspectives}

We show that CRs accelerated at SNR shocks can have an important impact on the plasma turbulence properties around SNRs. Slightly unexpectedly, self-excited Alfv\'en modes can provide some confinement of CRs around the sources in the dense cold ISM environments beyond 1 TeV due to a drop of the ion-neutral collision damping in the coupled regime. At lower energies, ion-neutral collisions are in the decoupled regime and strongly damp self-generated waves. However, a substantial number of waves can still be generated, especially if CRs are released at the end of the free expansion phase as in model $\mathcal{F}$. In that case, diffusion can be suppressed by two to three orders of magnitude over distances of a few tens of parsecs over a rime period of a few thousand years. These waves are driven by strong CR gradients which in turn can have some impact on ISM dynamics \citep{Commercon19}.

However, several assumptions made in our model limit its implications. First, we only consider the resonant streaming instability and slab Alfv\'en modes. We do not investigate other types of instabilities like the nonresonant streaming instability uncovered by \citet{2004MNRAS.353..550B}. Our model is only 1D and is therefore restricted to scales smaller than the coherence length of the background turbulence. One-dimensional geometry may also be problematic to investigate the inhomogeneous ISM, as is likely the case for atomic and molecular phases. The magnetic field lines can be mixed with turbulent motions in the super-Alfv\'enic regime which may occur for instance in the DiM phase, or possibly also for denser phases. A cylindrical 2D approach would be valuable to account for the geometry of the molecular clouds as well as the compression of the magnetic field lines \citep{2000ApJ...529..513C}. Two-dimensional calculations can also account for perpendicular CR diffusion. Another limit of our model is that it is restricted to medium where the speed of the scattering center remains approximately uniform. All these aspects, namely the geometry and adiabatic losses, will be accounted for in a future improved modeling. \\
Acknowledging these limitations, we argue that this simple model indicates that CR leakage from SNRs can induce a suppression of the diffusion with respect to what we could expect from the transport in the background turbulence. This effect should be taken into account in various aspects of the modeling of the ISM around CR sources.  

\begin{acknowledgements}
The authors thank S.Gabici, L. Nava, S. Recchia for fruitful discussions and for their comments on the manuscript. 
\end{acknowledgements}

\bibliographystyle{aa} 
\bibliography{Loann} 

\begin{appendix} 
\section{Cosmic ray propagation in the different ISM phases for the three models}
        \subsection{Results for the WNM}
    \begin{figure*} \centering
        \label{fig:pgd_WNM}
        \includegraphics[width=0.90\textwidth]{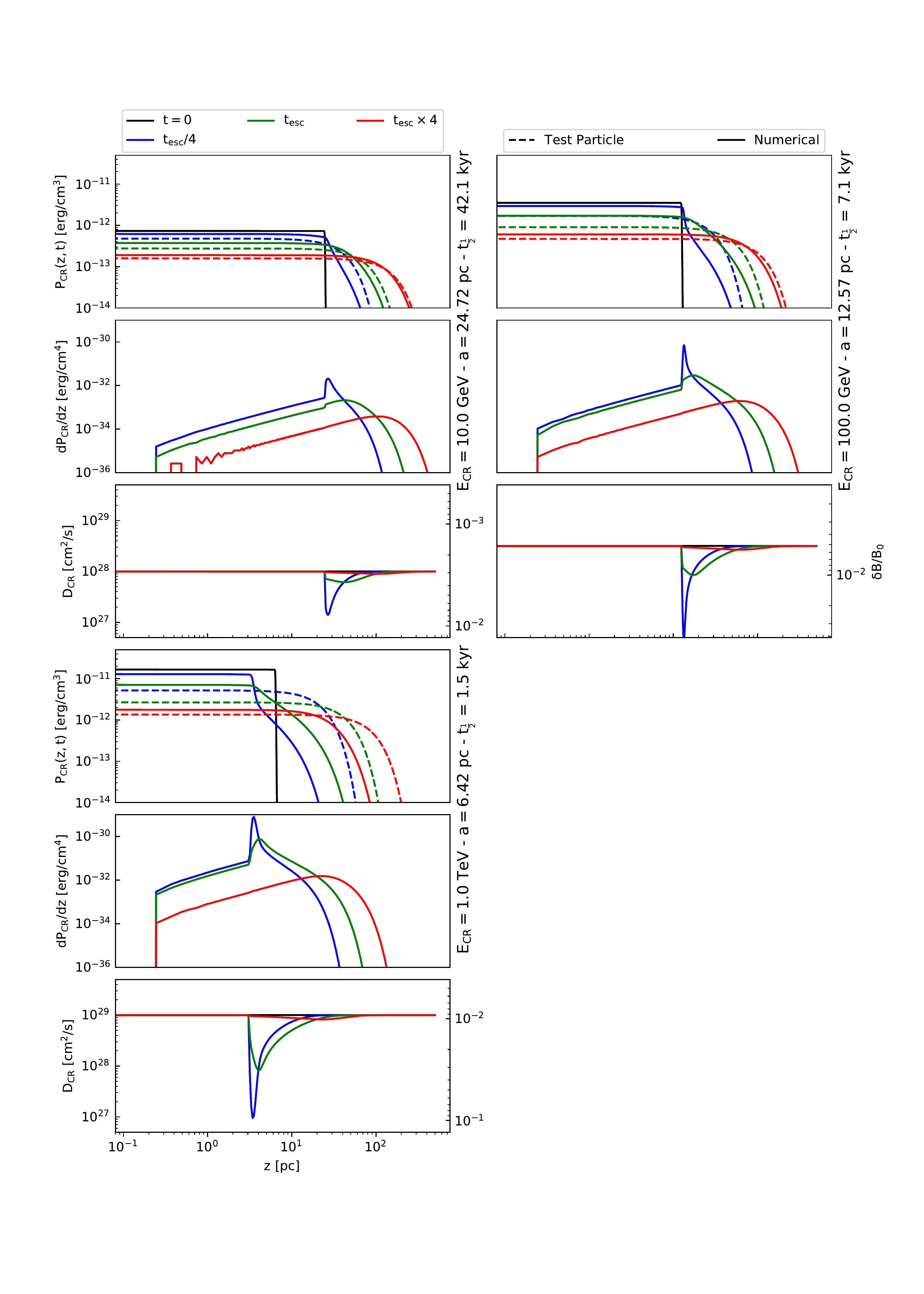}
        \caption{Time evolution of a CRC of initial radius $a$ in the WNM phase for model $\mathcal P$. The results at $10$ GeV, $100$ GeV, $1$ TeV, and $10$ TeV are shown from the left to the right and the top to the bottom. Each panel presents three plots and shows CR pressure evolution as a function of space for three times (top), the CR pressure gradient evolution (middle), and diffusion coefficients (bottom). Blue, green, and red  refer to $t_{1/2}/4$, $t_{1/2}$, and $4t_{1/2}$, respectively. The initial CR pressure distribution is represented by a thin black dashed line. The same typography is used for the background diffusion coefficient. Numerical solutions are represented with solid lines while test-particle solutions are presented by dashed lines. }
    \end{figure*}
    \newpage
    
     \begin{figure*} \centering
    \label{fig:pdg_WNM_2}
        \includegraphics[width=0.95\textwidth]{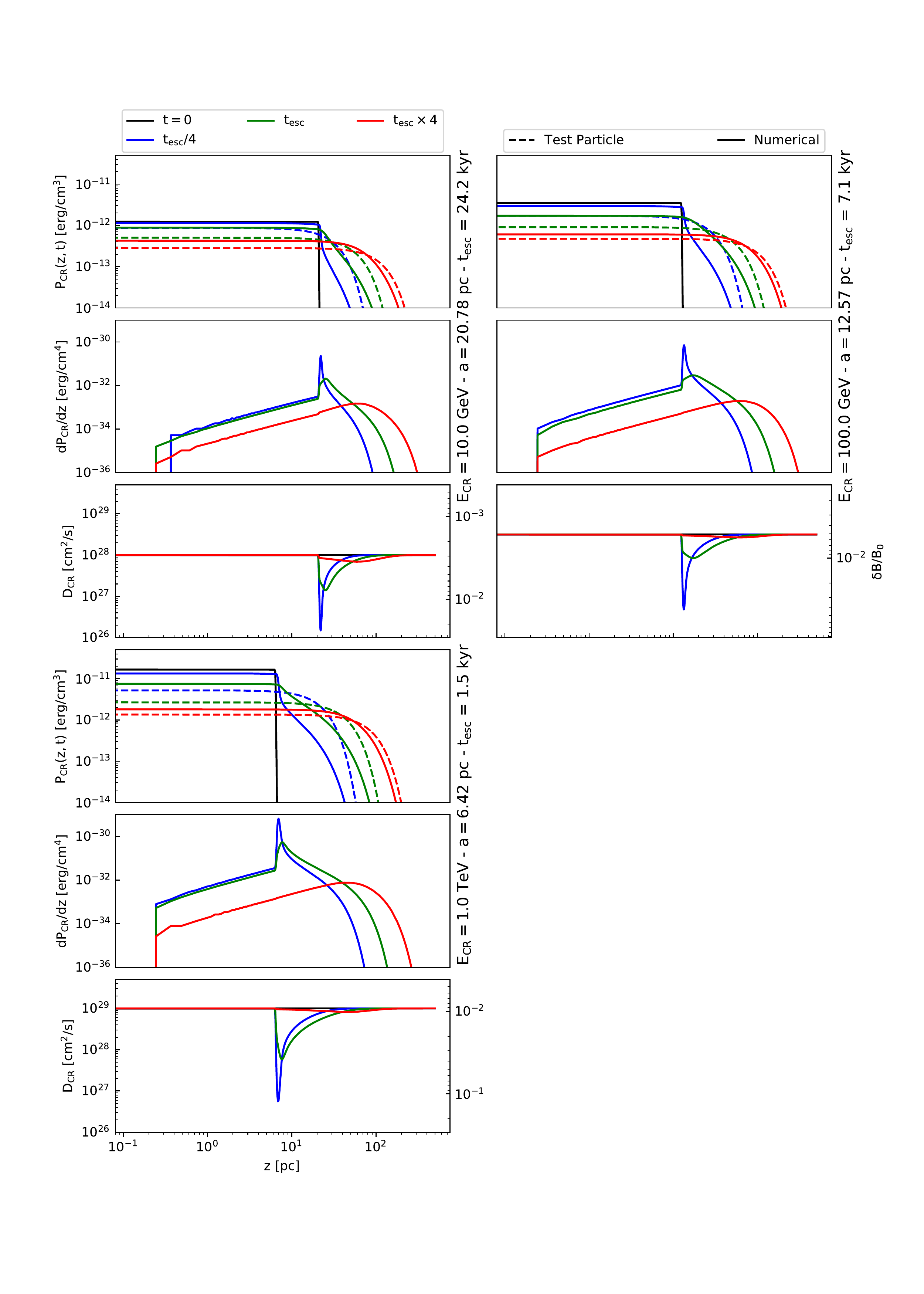}
        \caption{Time evolution of a CRC of initial radius $a$ in the WNM phase for model $\mathcal F$. Light gray plots show solutions differing from model $\mathcal P$. See figure (\ref{fig:pgd_WNM}) for more details.}
    \end{figure*}
    \newpage
     \begin{figure*} \centering
    \label{fig:pdg_WNM_3}
        \includegraphics[width=0.95\textwidth]{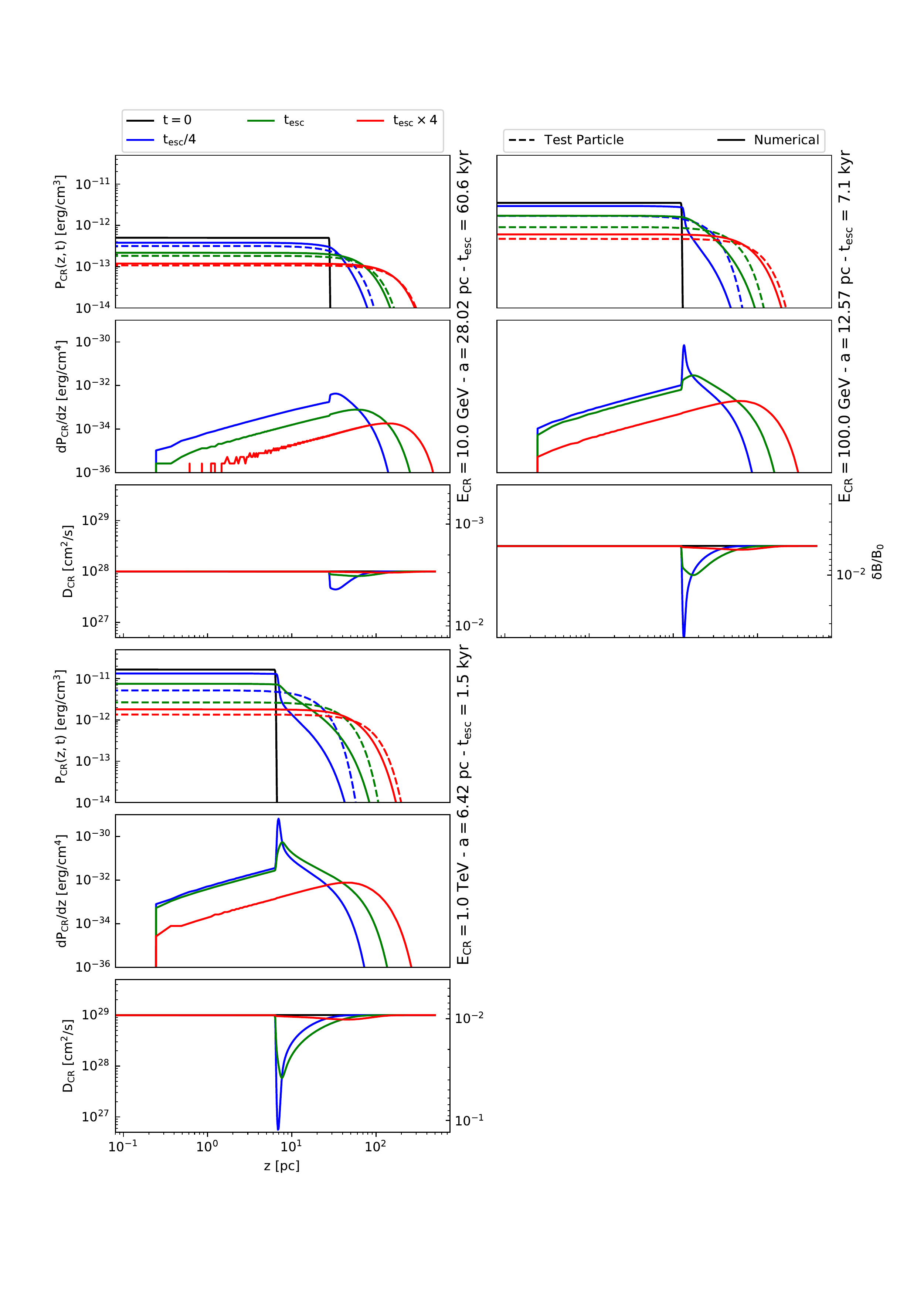}
        \caption{Time evolution of a CR cloud of initial radius $a$ in the WNM phase of the interstellar medium for the model $\mathcal S$. Light gray plots show solutions differing from model $\mathcal P$. See figure (\ref{fig:pgd_WNM}) for more details. }
    \end{figure*}
    \newpage
    
  \subsection{Results for the CNM}  
    \begin{figure*} \centering
    \label{fig:pdg_CNM}
        \includegraphics[width=0.95\textwidth]{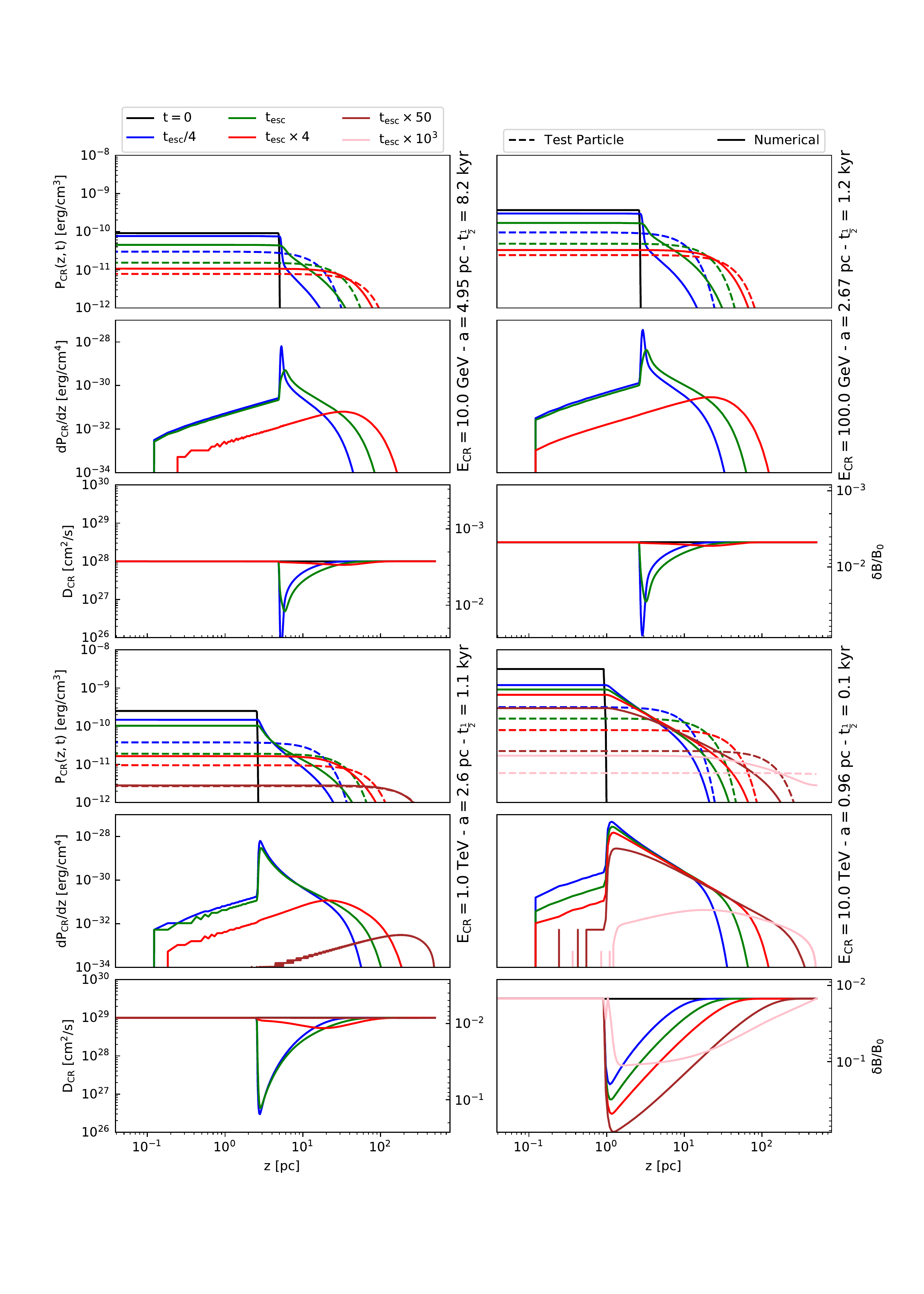}
        \caption{Time evolution of a CR cloud of initial radius $a$ in the CNM phase of the interstellar medium for model $\mathcal P$. See figure (\ref{fig:pgd_WNM}) for more details. In this medium we add two curves in brown and pink corresponding to solutions at 50 and $10^3$ $t_{1/2}$. }
    \end{figure*}
    \newpage
      \begin{figure*} \centering
    \label{fig:pdg_CNM_2}
        \includegraphics[width=0.95\textwidth]{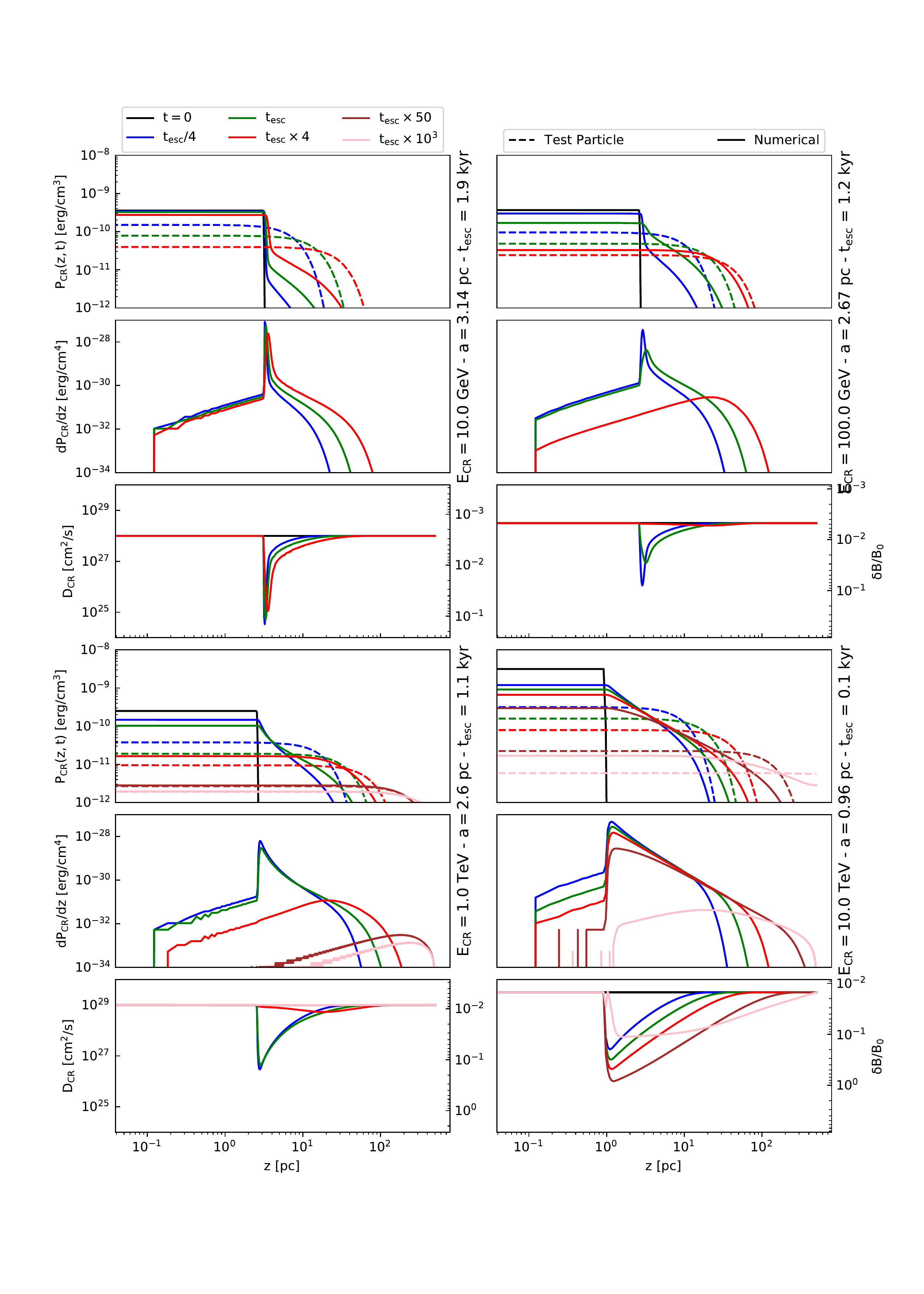}
        \caption{Time evolution of a CR cloud of initial radius $a$ in the CNM phase of the interstellar medium for model $\mathcal F$. Light gray plots refer to behaviors differing from model $\mathcal P$. See figure (\ref{fig:pgd_WNM}) for more details. }
    \end{figure*}
    \newpage
  \begin{figure*}\centering
    \label{fig:pdg_CNM_3}
        \includegraphics[width=0.95\textwidth]{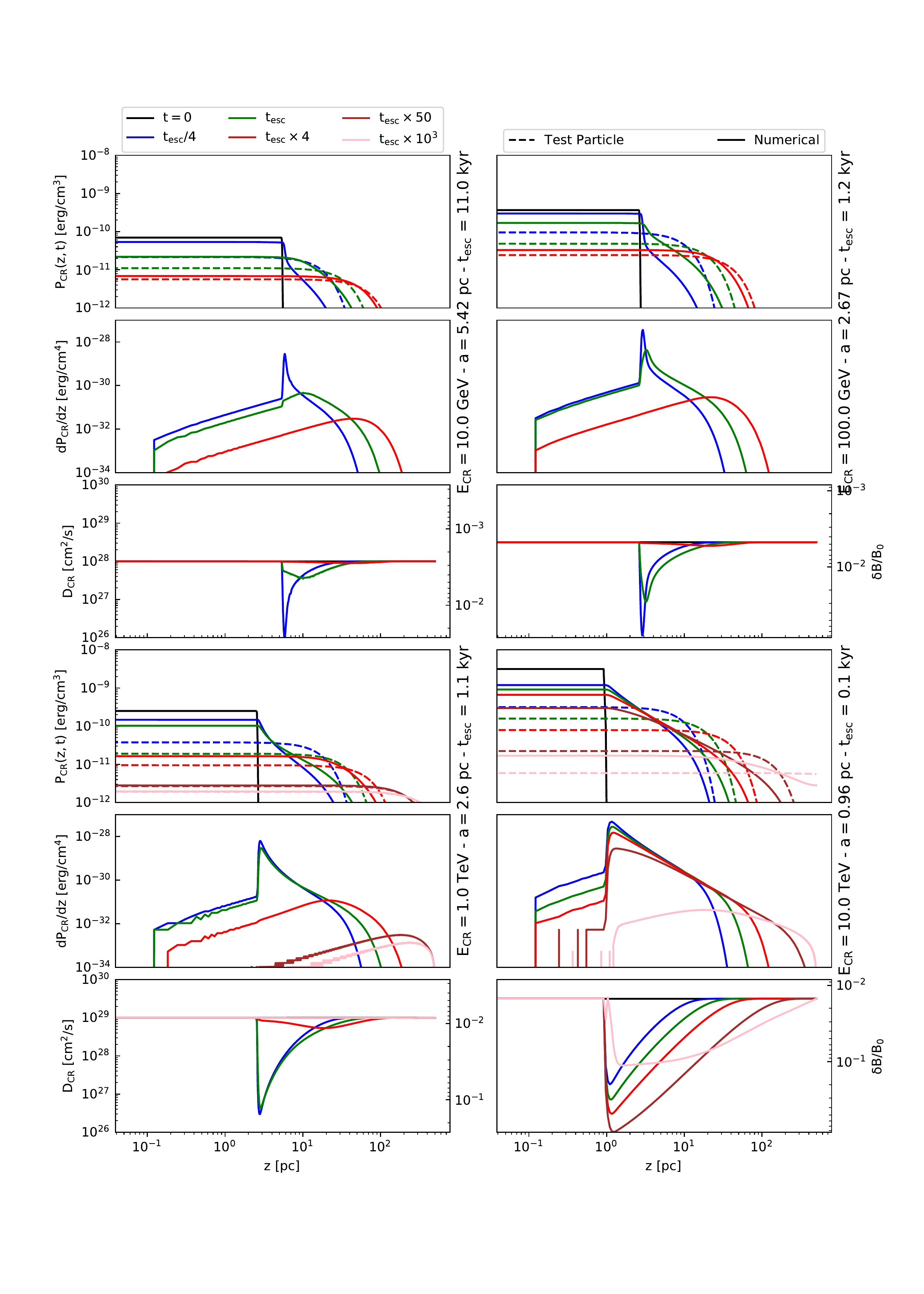}
        \caption{Time evolution of a CR cloud of initial radius $a$ in the CNM phase of the interstellar medium for model $\mathcal S$. Light gray plots refer to behaviors differing from model $\mathcal P$. See figure (\ref{fig:pgd_WNM}) for more details. }
    \end{figure*}
    \newpage
    
    \subsection{Results for the DiM medium}
    \begin{figure*} \centering
    \label{fig:pdg_DiM}
        \includegraphics[width=0.95\textwidth]{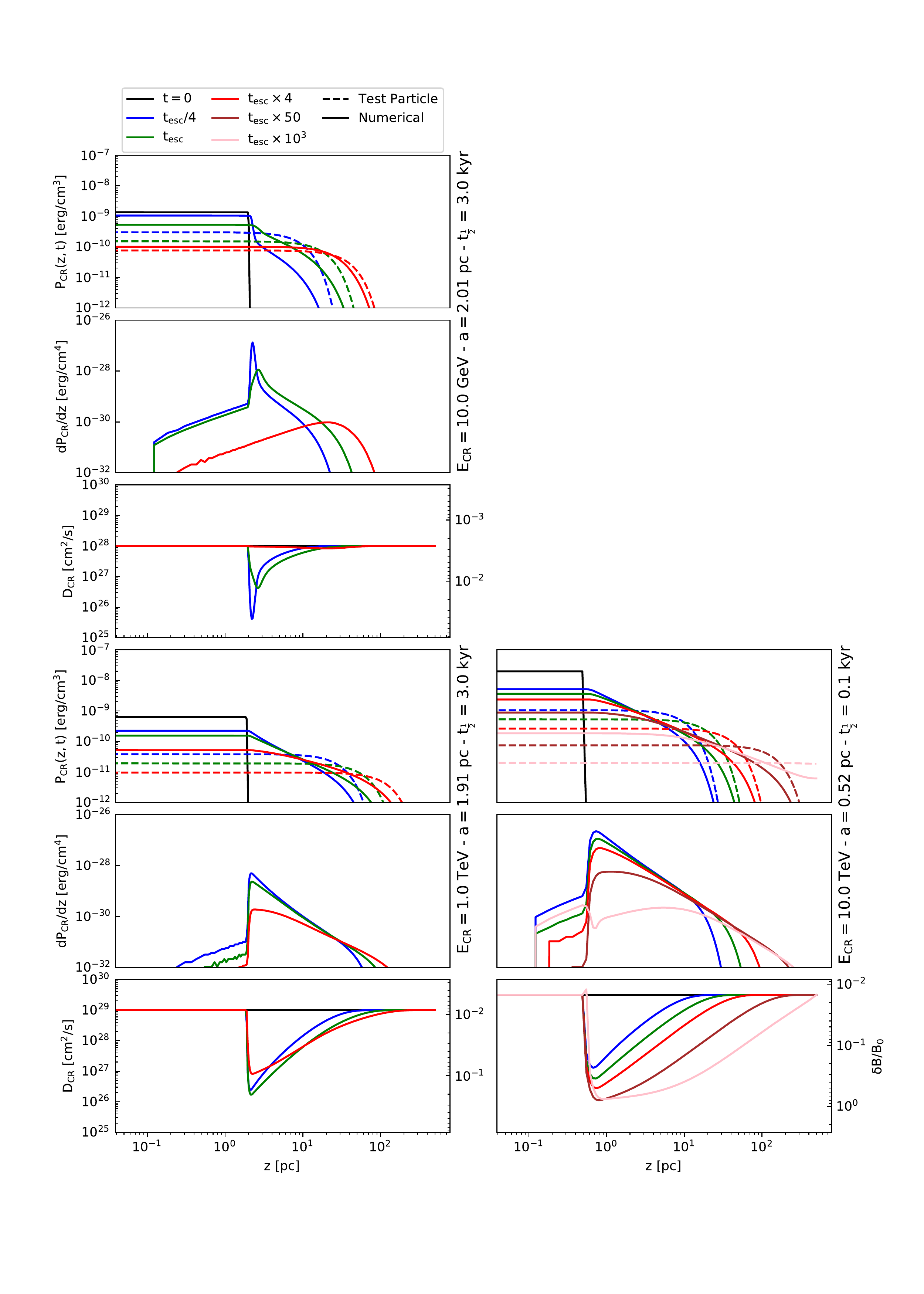}
        \caption{Time evolution of a CR cloud of initial radius $a$ in the DiM phase of the interstellar medium for model $\mathcal P$. See figure (\ref{fig:pgd_WNM}) for more details. In this medium we add two curves in brown and pink corresponding to solutions at 50 and $10^3$ $t_{1/2}$.}
    \end{figure*}
    \begin{figure*} \centering
    \label{fig:pdg_DiM_2}
        \includegraphics[width=0.95\textwidth]{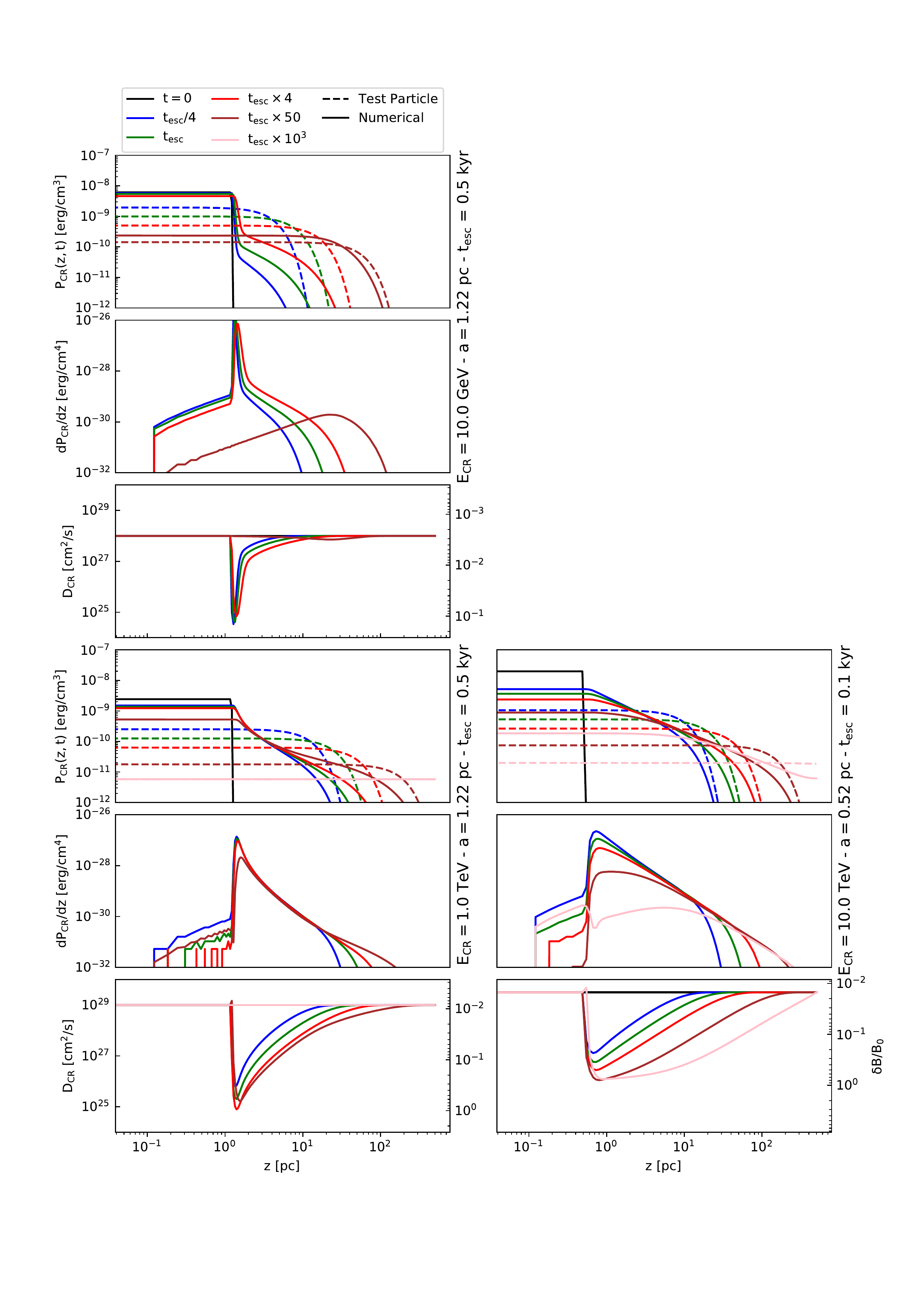}
        \caption{Time evolution of a CR cloud of initial radius $a$ in the DiM phase of the interstellar medium for model $\mathcal F$. Light gray plots refer to behaviors differing from model $\mathcal P$. See figure (\ref{fig:pgd_WNM}) for more details. }
    \end{figure*}
    \begin{figure*} \centering
    \label{fig:pdg_DiM_3}
        \includegraphics[width=0.95\textwidth]{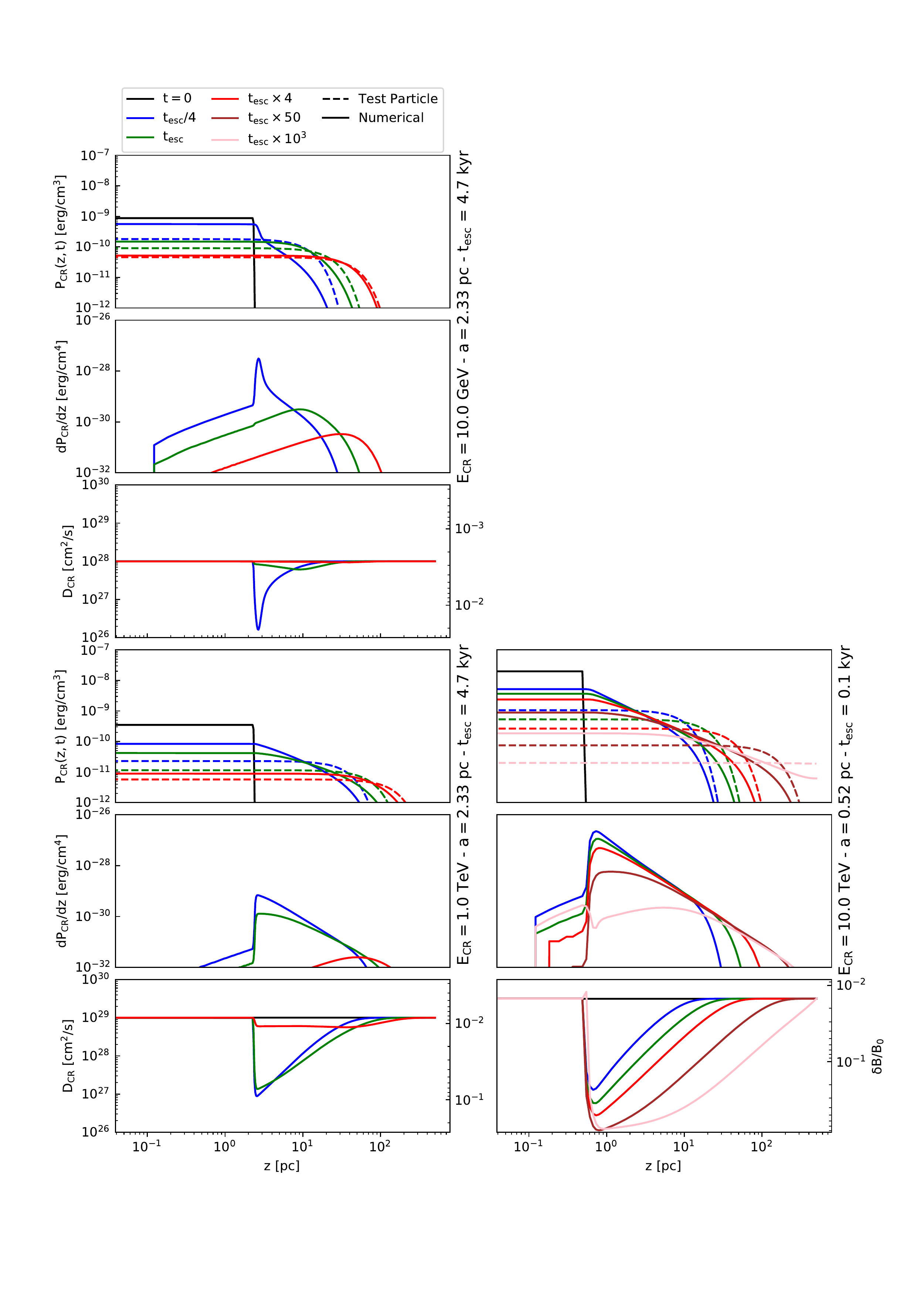}
        \caption{Time evolution of a CR cloud of initial radius $a$ in the DiM phase of the interstellar medium for model $\mathcal S$. Light gray plots refer to behaviors differing from model $\mathcal P$. See figure (\ref{fig:pgd_WNM}) for more details. }
    \end{figure*}
\end{appendix}
\end{document}